\documentclass[titlepage]{article}

\usepackage{makeidx}     
\usepackage{graphicx}    
\usepackage{multicol}    
\makeindex             
\usepackage{cite}
\let\mathbb\mathbf
\newcommand{\pr}{\mathop{\mathrm{pr}}\nolimits} 
\newcommand{\Sl}{\mathop{\mathrm{sl}}}
\newcommand{\SL}{\mathop{\mathrm{SL}}}
\newcommand{\const}{\mathop{\mathrm{const}}}

\newcommand{\semiprod}{\mathbin{\rlap{\raisebox{.1pt}{$\times$}}
\mskip-4mu\supset}}

\newcommand{\semisum}{\mathbin{\rlap{\raisebox{.3pt}{\small $+$}}
\mskip-4mu\supset}} 

\newcommand{\arctanh}{\mathop{\mathrm{arctanh}}\nolimits}

\usepackage{amsthm}

\newtheorem{theorem}{Theorem}

\theoremstyle{definition}
\newtheorem{example}{Example}
\newtheorem{definition}{Definition}

4
\begin{document}

\title{Symmetries of Discrete Systems}

\author{Pavel Winternitz\\
Centre de recherches math\'ematiques and\\ 
D\'epartement de math\'ematiques et de statistique\\
Universit\'e de Montr\'eal\\
C.P. 6128, succ.\ Centre-Ville\\
Montr\'eal, QC H3C 3J7\\
 Canada\\
\texttt{wintern@crm.umontreal.ca}}

\date{CRM-2932\\
September 2003}

\maketitle

\begin{abstract}
In this series of lectures, presented at the CIMPA Winter School on Discrete Integrable Systems in February 2003, we give a review of the application of Lie point symmetries, and their generalizations, to the study of difference equations. The overall theme could be called ``continuous symmetries of discrete equations''.
\end{abstract}

\tableofcontents

\section{Introduction}\label{sec1}

\subsection{Symmetries of Differential Equations}\label{subsec1.1}

 Before studying the symmetries of difference equations, let us very briefly
review the theory of the symmetries of differential equations. For all
details, proofs and further information we refer to the many excellent books 
on the subject e.g.\ \cite{ref1,ref2,ref3,ref4,ref5,ref6,ref7,ref8}.

Let us consider a completely general system of differential equations
\begin{equation}\label{1.1}
E_a(x, u, u_x, u_{xx}, \dots u_{nx}) = 0, \quad
x \in \mathbb R^p, u \in \mathbb R^q, a = 1, \dots N,
\end{equation}
where e.g. $u_{nx}$ denotes all (partial) derivatives of order $n$. The numbers
$p, q, n$ and $N$ are all nonnegative integers.

We are interested in the symmetry group $G$ of the system (\ref{1.1}), i.e. in
the local Lie group of local point transformations taking solutions of
eq.~(\ref{1.1}) into solutions. Point transformations in the space $X \times U$
of independent and dependent variables have the form
\begin{equation}\label{1.2}
\tilde x = \Lambda_{\lambda}(x, u), \quad \tilde u = \Omega_{\lambda}(x, u),
\end{equation}
where $\lambda$ denotes the group parameters. We have
\[
\Lambda_0(x, u) = x, \quad \Omega_0(x, u) = u\nonumber
\]
and the inverse transformation $(\tilde x, \tilde u) \to (x, u)$ exists, at
least locally.

The transformations (\ref{1.2}) of local coordinates in  $X \times U$ also
determine the transformations of functions $u = f(x)$ and of derivatives of
functions. A group $G$ of local point transformations of $X \times U$ will be a
symmetry group of the system (\ref{1.1}) if the fact that $u(x)$ is a solution
implies that $\tilde u(\tilde x)$ is also a solution.

The two fundamental questions to ask are:
\begin{enumerate}
\item[1.] How to find the maximal symmetry group $G$ for a given system of
equations (\ref{1.1})?

\item[2.] Once the group $G$ is found, what do we do with it?
\end{enumerate}

Let us first discuss the question of motivation. The symmetry group $G$ allows
us to do the following.

\begin{enumerate}
\item[1.] Generate new solutions from known ones. Sometimes trivial solutions
can be boosted into interesting ones.

\item[2.] Identify equations with isomorphic symmetry groups. Such equations
may be transformable into each other. Sometimes nonlinear equations can be
transformed into linear ones.

\item[3.] Perform symmetry reduction: reduce the number of variables in a PDE
and obtain particular solutions, satisfying particular boundary conditions:
group invariant solutions. For ODEs of order $n$, we can reduce the order of the
equation. In this reduction, there is no loss of information. If we can reduce
the order to zero, we obtain a general solution depending on $n$ constants, or a
general integral (an algebraic equation depending on $n$ constants).
\end{enumerate}

How does one find the symmetry group $G$? One looks for infinitesimal
transformations, i.e.\ one looks for the Lie algebra $L$ that corresponds to
$G$. Instead of looking for ``global'' transformations as in eq.~(\ref{1.2}) one
looks for infinitesimal ones. A one-parameter group of infinitesimal point
transformations will have the form
\begin{eqnarray}
\tilde x_i & = & x_i + \lambda\xi_i(x, u) \quad |\lambda| << 1\label{1.3}\\
\tilde u_{\alpha} & = & u_{\alpha} + \lambda\phi_{\alpha}(x, u) \quad 1 \leq i
\leq p, \quad 1 \leq \alpha \leq q.\nonumber
\end{eqnarray}
The functions $\xi_i$ and $\phi_{\alpha}$ must be found from the condition that
$\tilde u(\tilde x)$ is a solution whenever $u(x)$ is one. The derivatives
$\tilde u_{\alpha, \tilde x_i}$ must be calculated using eq.~(\ref{1.3}) and
will involve derivatives of $\xi_i$ and $\phi_{\alpha}$. A $K$-th derivative of
$\tilde u_{\alpha}$ with respect to the variable $\tilde x_i$ will involve
derivatives of $\xi_i$ and $\phi_{\alpha}$ up to order $K$. We then substitute
the transformed quantities into eq.~(\ref{1.1}) and request that the equation be
satisfied for $\tilde u(\tilde x)$, whenever it is satisfied for $u(x)$. Thus,
terms of order $\lambda^0$ will drop out. Terms of order $\lambda$ will
provide a system of determining equations for $\xi_i$ and $\phi_{\alpha}$.
Terms of order $\lambda^k$, $k = 2, 3, \dots$ are to be ignored, since we are
looking for infinitesimal symmetries.

The functions $\xi_i$ and $\phi_{\alpha}$ depend only on $x$ and $u$, not on
first, or higher derivatives, $u_{\alpha, x_i}$, $u_{\alpha, x_ix_k}$, etc. This
is actually the definition of ``point'' symmetries. The determining equations
will explicitly involve derivatives of $u_{\alpha}$, up to the order $n$ (the
order of the studied equation). The coefficients of all linearly independent
expressions in the derivatives must vanish separately. This provides a system
of determining equations for the functions $\xi_i(x, u)$ and $\phi_{\alpha}(x,
u)$. This is a system of linear partial differential equations of order $n$. The
determining equations are linear, even if the original system (\ref{1.1}) is
nonlinear. This ``linearization'' is due to the fact that all terms of order
$\lambda^j$, $j \geq 2$, are ignored.

The system of determining equations is usually overdetermined, i.e.\ there are
usually more determining equations than unknown functions $\xi_i$ and
$\phi_{\alpha}$ ($p+q$ functions). The independent variables in the determining
equations are $x \in \mathbb R^p$, $u \in \mathbb R^q$.

For an overdetermined system, three possibilities occur.

\begin{enumerate}
\item[1.] The only solution is the trivial one $\xi_i = 0$, $\phi_{\alpha} =
0$,  $i = 1, \dots p, \alpha = 1, \dots, q$. In this case the symmetry algebra
is $L = \{0\}$, the symmetry group is $G = I$ and the symmetry method is to no
avail.

\item[2.] The general solution of the determining equations depends on a finite
number $K$ of constants. In this case the studied system (\ref{1.1}) has a
finite-dimensional Lie point symmetry group and we have $\dim G = K$.

\item[3.] The general solution depends on a finite number of arbitrary
functions of some of the variables $\{x_i, u_{\alpha}\}$. In this case the
symmetry group is infinite dimensional. This last case is of particular
interest.
\end{enumerate}

The search for the symmetry algebra $L$ of a system of differential equations
is best formulated in terms of vector fields acting on the space $X \times U$
of independent and dependent variables. Indeed, consider the vector field
\begin{equation}\label{1.4}
X = \sum^p_{i=1}\xi_i(x, u)\partial x_i + \sum^q_{\alpha=1}\phi_{\alpha}
(x, u)\partial u_{\alpha},
\end{equation}
where the coefficients $\xi_i$ and $\phi_{\alpha}$ are the same as in
eq.~(\ref{1.3}). If these functions are known, the vector field (\ref{1.4}) can
be integrated to obtain the finite transformations (\ref{1.2}). Indeed, all we
have to do is integrate the equations
\begin{equation}\label{1.5}
\frac{d\tilde x_i}{d\lambda} = \xi_i(\tilde x, \tilde u), \quad \frac{d\tilde
u_{\alpha}}{d\lambda} = \phi_{\alpha}(\tilde x, \tilde u),
\end{equation}
subject to the initial conditions
\begin{equation}\label{1.6}
\tilde x_i\mid_{\lambda = 0} = x_i \quad \tilde u_{\alpha}\mid_{\lambda=0} =
u_{\alpha}.
\end{equation}
This provides us with a one-parameter group of local Lie point
transformations of the form (\ref{1.2}) with $\lambda$ the group parameter.

The vector field (\ref{1.4}) tells us how the variables $x$ and $u$ transform.
We also need to know how derivatives like $u_x$, $u_{xx}$, $\dots$ transform.
This is given by the prolongation of the vector field $X$.

We have
\begin{eqnarray}
\lefteqn{\pr X  =  X +
\sum_{\alpha}\biggl\{\sum_i\phi^{x_i}_{\alpha}\partial u_{x_i} +\sum_{i,
k}\phi^{x_ix_k}_{\alpha}\partial u_{x_ix_k}}\label{1.7}\\ 
&\hspace{2in} & +  \sum^{x_ix_kx_l}_{i, k, l}\phi_{\alpha}\partial u_{x_ix_kx_l}
+
\dots\biggr\},\nonumber
\end{eqnarray}
where the coefficients in the prolongation can be calculated recursively, using
the total derivative operator
\begin{equation}\label{1.8}
D_{x_i} = \partial_{x_i}+ u_{\alpha, x_i}\partial_{u_{\alpha}}+u_{\alpha,
x_ax_i}\partial_{u_{\alpha}, x_a} + u_{\alpha, x_ax_bx_i}\partial_{u_{\alpha},
x_ax_b} + \dots
\end{equation}
(a summation over repeated indices is to be understood).

The recursive formulas are
\begin{eqnarray}
\phi^{x_i}_{\alpha} & = & D_{x_i}\phi_{\alpha} - (D_{x_i}\xi_a)u_{\alpha, x_a}
\nonumber\\
\phi^{x_ix_k}_{\alpha} & = & D_{x_k}\phi^{x_i}_{\alpha} - (D_{x_k}\xi_a)
u_{\alpha, x_ix_a}\label{1.9}\\
\phi^{x_ix_kx_l}_{\alpha} & = & D_{x_l}\phi^{x_ix_k}_{\alpha} - (D_{x_l}\xi_a)
u_{\alpha, x_ix_kx_a}\nonumber
\end{eqnarray}
etc.

The $n$-th prolongation of $\widehat X$ acts on functions of $x$, $u$ and
all derivatives of $u$ up to order $n$. It also tells us how
derivatives transform. Thus, to obtain the transformed quantities
$\tilde u_{\tilde x_i}$ we must integrate eq. (\ref{1.5}) with conditions
(\ref{1.6}), together with
\begin{equation}\label{1.10}
\frac{d\tilde u_{\tilde x_i}}{d\lambda} = \phi^{x_i}(\tilde x, \tilde u, \tilde
u_{\tilde x}), \quad \tilde u_{\tilde x}\mid_{\lambda=0} = u_x.
\end{equation}
We see that the coefficients of the prolonged vector field are
expressed in terms of derivatives of $\xi_i$ and $\phi_{\alpha}$,
the coefficients of the original vector field. They carry no new
information: the transformation of derivatives is completely
determined, once the transformations of functions are known.

The invariance condition for the system (\ref{1.1}) is expressed in
terms of the operator (\ref{1.7}) as
\begin{equation}\label{1.11}
\pr^{(n)}XE_a\mid_{E_1=\dots = E_N = 0} = 0, \quad a = 1, \dots N,
\end{equation}
where $\pr^{(n)}X$ is the prolongation (\ref{1.7}) calculated up to
order $n$ where $n$ is the order of the system (\ref{1.1}).

In practice the symmetry algorithm consists of several steps, most
of which can be carried out on a computer. For early computer
programs calculating symmetry algebras, see Ref.~\cite{ref9,ref10}.
For a more recent review, see \cite{ref11}.

The individual steps are:
\begin{enumerate}
\item[1.] Calculate all the coefficients in the $n$-th prolongation of
$\widehat X$. This depends only on the order of the system (\ref{1.1}),  i.e.
$n$, and on the number of independent and dependent variables, i.e. $p$ and $q$.

\item[2.] Consider the system (\ref{1.1}) as a system of algebraic equations for
$x$, $u$, $u_x$, $u_{xx}$, etc. Choose $N$ variables $v_1$, $v_2$, $\dots v_N$
and solve the system (\ref{1.1}) for these variables. The $v_i$ must satisfy the
following conditions.
\begin{enumerate}
\item[(i)] Each $v_i$ is a derivative of one $u_{\alpha}$ of at least order 1.
\item[(ii)] The variables $v_i$ are all independent, none of them is a
derivative of any other one.
\item[iii)] No derivatives of any of the $v_i$ figure in the system (\ref{1.1}).
\end{enumerate}

\item[3.] Apply $\pr^{(n)}X$ to all the equations in (\ref{1.1}) and eliminate
all expressions $v_i$ from the result. This provides us with the system
(\ref{1.11}).

\item[4.] Determine all linearly independent expressions in the derivatives
remaining in (\ref{1.11}), once the quantities $v_i$ are eliminated. Set the
coefficients of these expressions equal to zero. This provides us with the
determining equations, a system of linear partial differential equations of
order $n$ for $\phi_{\alpha}(x, u)$ and $\xi_i(x, u)$.

\item[5.] Solve the determining equations to obtain the symmetry algebra.

\item[6.] Integrate the obtained vector fields to obtain the one-parameter
subgroups of the symmetry group. Compose them appropriately to obtain the
connected component $G_o$ of the symmetry group $G$.

\item[7.] Extend the connected component $G_o$ to the full group $G$ by adding
all discrete transformations leaving the system (\ref{1.1}) invariant. These
discrete transformations will form a finite, or discrete group $G_D$. We have
\begin{equation}\label{1.12}
G = G_D \semiprod G_o
\end{equation}
i.e. $G_o$ is an invariant subgroup of $G$.
\end{enumerate}

Let us consider the case when at Step 5 we obtain a finite
dimensional Lie algebra $L$, i.e. a vector field $X$ depending on $K$
parameters, $K \in \mathbb Z^>$, $K < \infty$. We can then choose a basis
\begin{equation}\label{1.13}
\{X_1, X_2, \dots, X_K\}
\end{equation} 
of the Lie algebra $L$. The basis that is naturally obtained in this
manner depends on our integration procedure, though the algebra $L$
itself does not. It is useful to transform the basis (\ref{1.13}) to a
canonical form in which all basis independent properties of $L$ are
manifest. Thus, if $L$ can be decomposed into a direct sum of
indecomposable components,
\begin{equation}\label{1.14}
L = L_1 \oplus L_2 \oplus \dots \oplus L_M,
\end{equation}
then a basis should be chosen that respects this decomposition. The
components $L_i$ that are simple should be identified according to the
Cartan classification (over $\mathbb C$) or the Gantmakher
classification (over $\mathbb R$) \cite{ref12,ref13}. The components
that are solvable should be so organized that their nilradical
\cite{ref14,ref15} is manifest. For those components that are neither
simple, nor solvable, the basis should be chosen so as to respect the
Levi decomposition \cite{ref14,ref15}.         

So far we have considered only point transformations, as in eq.~(\ref{1.2}), in
which the new variables $\tilde x$ and $\tilde u$ depend only on the old ones,
$x$ and $u$. More general transformations are ``contact transformations'',
where $\tilde x$ and $\tilde u$ also depend on first derivatives of $u$. A
still more general class of transformations are generalized transformations,
also called ``Lie-B\"acklund'' transformations \cite{ref1,ref16}. For these
we have
\begin{eqnarray}
\tilde x & = & \Lambda_{\lambda}(x, u, u_x, u_{xx}, \dots)\\
\tilde u & = & \Omega_{\lambda}(x, u, u_x, u_{xx}, \dots)\nonumber
\end{eqnarray}
involving derivatives up to an arbitrary order. The coefficients $\xi_i$ and
$\phi_{\alpha}$ of the vector fields (\ref{1.4}) will then also depend on
derivatives of $u_{\alpha}$.

When studying generalized symmetries, and sometimes also point symmetries, it
is convenient to use a different formalism, namely that of evolutionary vector
fields.

Let us first consider the case of Lie point symmetries, i.e.\ vector fields of
the form (\ref{1.4}) and their prolongations (\ref{1.7}). To each vector field
(\ref{1.4}) we can associate its evolutionary counterpart $X_e$, defined as
\begin{eqnarray}
X_e & = & Q_{\alpha}(x, u, u_x)\partial u_{\alpha},\label{1.16}\\
Q_{\alpha} & = & \phi_{\alpha} - \xi_i\frac{\partial  u_{\alpha}}{\partial
x_i}.\label{1.17}
\end{eqnarray}
The prolongation of the evolutionary vector field (\ref{1.16}) is defined as
\begin{eqnarray}
\pr X_e &=& Q_{\alpha}\partial u_a + Q^{x_i}_{\alpha} \partial u_{\alpha, x_i} +
Q^{x_ix_k}_{\alpha}\partial u_{\alpha, x_ix_k} + \dots\label{1.18}\\
Q^{x_i}_{\alpha} &=& D_{x_i}Q_{\alpha}, \quad Q^{x_ix_k}_{\alpha} = D_{x_i}D_
{x_k}Q_{\alpha}, \dots\nonumber.
\end{eqnarray}
The functions $Q_{\alpha}$ are called the characteristics of the vector field.
Notice that $X_e$ and $\pr X_e$ do not act on the independent variables $x_i$.

For Lie point symmetries evolutionary and ordinary vector fields are entirely
equivalent and it is easy to pass from one to the other. Indeed,
eq.~(\ref{1.17}) gives the connection between the two.

The symmetry algorithm for calculating the symmetry algebra $L$ in terms of
evolutionary vector fields is also equivalent. Eq.~(\ref{1.11}) is simply
replaced by
\begin{equation}\label{1.19}
\pr^{(n)}X_e E_a\mid_{E_1=\dots=E_N=0}, = 0, \quad a = 1, \dots N.
\end{equation}
The reason that eq.~(\ref{1.11}) and (\ref{1.19}) are equivalent is the
following. It is easy to check that we have
\begin{equation}\label{1.20}
\pr^{(n)}X_e = \pr^{(n)}X - \xi_iD_i.
\end{equation}

The total derivative $D_i$ is itself a generalized symmetry of eq.~(\ref{1.1}),
i.e. we have
\begin{equation}\label{1.21a}
D_iE_a\mid_{E_1=E_2=\dots=E_n=0}, =0 \quad i = 1, \dots p, \quad a = 1, \dots N.
\end{equation}
Eq.~(\ref{1.20}) and (\ref{1.21a}) prove that the systems (\ref{1.11}) and
(\ref{1.19}) are equivalent. Eq.~(\ref{1.21a}) itself follows from the fact that
$DE_a=0$ is a differential consequence of eq.~(\ref{1.1}), hence every
solution of eq.~(\ref{1.1}) is also a solution of eq.~(\ref{1.21a}).

To find generalized symmetries of order $k$ we use eq.~(\ref{1.16}) but allow
the characteristics $Q_{\alpha}$ to depend on all derivatives  of $u_{\alpha}$
up to order $k$. The prolongation is calculated using eq.~(\ref{1.18}). The
symmetry algorithm is again eq.~(\ref{1.19}).

A very useful property of evolutionary symmetries is that they provide
compatible flows. This means that the system of equations
\begin{equation}\label{1.21b}
\frac{\partial u_{\alpha}}{\partial\lambda} = Q_{\alpha}
\end{equation}
is compatible with the system (\ref{1.1}). In particular, group invariant
solutions, i.e.\ solutions invariant under a subgroup of $G$ are obtained as
fixed points
\begin{equation}\label{1.22}
Q_{\alpha} = 0.
\end{equation}
If $Q_{\alpha}$ is the characteristic of a point transformation then
(\ref{1.22}) is a system of quasilinear first order partial differential
equations. They can be solved, the solution substituted into (\ref{1.1}) and
this provides the invariant solutions explicitly.

\subsection{Comments on Symmetries of Difference Equations}\label{subsec1.2}

The study of symmetries of difference equations is much more recent than that
of differential equations. Early work in this direction is due to Maeda
\cite{ref17,ref18} who mainly studied transformations acting on the dependent
variables only. A more recent series of papers was devoted to Lie point
symmetries of differential-difference equations on fixed regular lattices
\cite{ref19,ref20,ref21,ref22,ref23,ref24,ref25,ref26,ref27,ref28}. A different
approach was developed mainly for linear or linearizable difference equations
and involved transformations acting on more than one point of the lattice
\cite{ref29,ref30,ref31,ref32,ref33}. The symmetries considered in this
approach are really generalized ones, however they reduce to point ones in the
continuous limit.

A more general class of generalized symmetries has also been investigated for
difference equations, and differential-difference equations on fixed regular
lattices \cite{ref34,ref35,ref36,ref37}.

A different approach to symmetries of discrete equations was originally
suggested by V. Dorodnitsyn and collaborators \cite{ref38,ref39,ref40,
ref41,ref42,ref43,ref44,ref45,ref46,ref47}. The main aim of this series of
papers is to discretize differential equations while preserving their Lie point
symmetries.

Symmetries of ordinary and partial difference schemes on lattices that are a
priori given, but are allowed to transform under point transformations, were
studied in Ref.~\cite{ref48,ref49,ref50}.

\section{Ordinary Difference Schemes and Their Point Symmetries}\label{sec2}

\subsection{Ordinary Difference Schemes}\label{subsec2.1}

An ordinary differential equation (ODE) of order $n$ is a relation involving
one independent variable $x$, one dependent variable $u = u(x)$ and $n$
derivatives $\stackrel{\prime}{u}$, $\stackrel{\prime\prime}{u}$, $\dots
u^{(n)}$
\begin{equation}\label{2.1}
E(x, u, \stackrel{\prime}{u}, \stackrel{\prime\prime}{u}, \dots, u^{(n)}) = 0
\quad \frac{\partial E}{\partial u^{(n)}} \neq 0. 
\end{equation}

An ordinary difference scheme (O$\Delta$S) involves two objects, a difference
equation and a lattice. We shall specify an O$\Delta$S by a system of two
equations, both involving two continuous variables $x$ and $u(x)$, evaluated at
a discrete set of points $\{x_n\}$.

Thus, a difference scheme of order $K$ will have the form
\begin{equation}\label{2.2}
\begin{array}{c}
E_a(\{x_k\}^{n+N}_{k=n+M}, \{u_k\}^{n+N}_{k=n+M}) = 0, a = 1, 2\\[\jot]
K= N  - M + 1, \quad n, M, N \in \mathbb Z, \quad u_k \equiv u(x_k).
\end{array}
\end{equation}
At this stage we are not imposing any boundary conditions, so the reference
point $x_n$ can be arbitrarily shifted to the left, or to the right. The order
$K$ of the system is the number of points involved in the scheme (\ref{2.2})
and it is assumed to be finite. We also assume that if the values of $x_k$ and
$u_k$ are specified in $(N-M)$ neighbouring point, we can calculate their
values in the point to the right, or to the left of the given set, using
equations (\ref{2.2}).

A continuous limit for the spacings between all neighbouring points going to
zero, if it exists, will take one of the equations (\ref{2.2}) into a
differential equation of order $K' \leq K$, another into an identity (like $0 =
0$).

When taking the continuous limit it is convenient to introduce different
quantities, namely differences between neighbouring points and discrete
derivatives like
\begin{eqnarray}
h_+ (x_n) & = & x_{n+1}-x_n, \quad h_-(x_n) =x_n - x_{n-1},\nonumber\\
u_{,x} & = & \frac{u_{n+1}-u_n}{x_{n+1}-x_n}, \quad u_{,\underline 
x}= \frac{u_n-u_{n-1}}{x_n-x_{n-1}},\label{2.3}\\
u_{,x\underline x} & = &
2\frac{u_{,x}-u_{,\underline x}}{x_{n+1}-x_{n-1}},\dots\nonumber
\end{eqnarray}

In the continuous limit, we have
\[
h_+ \to 0, \quad h_- \to 0, \quad u_{,{x}} \to \stackrel{\prime}{u},\quad
u_{,{\underline x}} \to \stackrel{\prime}{u}, \quad u_{,x\underline x} \to
\stackrel{\prime\prime}{u}.
\]

As a clarifying example of the meaning of the difference scheme (\ref{2.2}),
let us consider a three point scheme that will approximate a second order
linear difference equation:
\begin{eqnarray}
E_1 & = & \frac{u_{n+1}-2u_n+u_{n-1}}{(x_{n+1}-x_n)^2} - u_n = 0,\label{2.4}\\
E_2 & = & x_{n+1} - 2x_n + x_{n-1} = 0.\label{2.5} 
\end{eqnarray}
The solution of eq. $E_2=0$, determines a uniform lattice
\begin{equation}\label{2.6}
x_n = hn + x_0
\end{equation}
The scale $h$ and the origin $x_0$ in eq.~(\ref{2.6}) are not fixed by
eq.~(\ref{2.5}), instead they appear as integration constants, i.e.\ they are a
priori arbitrary. Once they are chosen, eq.~(\ref{2.4}) reduces to a linear
difference equation with constant coefficients, since we have $x_{n+1} - x_n =
h$. Thus, a solution of eq.~(\ref{2.4}) will have the form
\begin{equation}\label{2.7}
u_n = \lambda^{x_n}.
\end{equation}
Substituting (\ref{2.7}) into (\ref{2.4}) we obtain the general solution of the
difference scheme (\ref{2.4}), (\ref{2.5}) as
\begin{eqnarray}
u(x_n) & = & c_1\lambda^{x_n}_1+c_2\lambda^{x_n}_2, \quad x_n = hn +
x_0,\label{2.8}\\
\lambda_{1, 2} & = & \left(\frac{2+h^2\pm h\sqrt{4+h^2}}{2}\right)^{1/2}.
\nonumber
\end{eqnarray}
The solution (\ref{2.8}) of the system (\ref{2.4}), (\ref{2.5}) depends on 4
arbitrary constants $c_1$, $c_2$, $h$ and $x_0$.

Now let us consider a general three point scheme of the form
\begin{equation}\label{2.9}
E_a(x_{n-1}, x_n, x_{n+1}, u_{n-1}, u_n, u_{n+1}) = 0, \quad a = 1, 2
\end{equation}
satisfying
\begin{equation}\label{2.10}
\det\left(\frac{\partial(E_1, E_2)}{\partial(x_{n+1}, u_{n+1})}\right) \neq 0,
\quad \det\left(\frac{\partial(E_1, E_2)}{\partial(x_{n-1}, u_{n-1})}\right)
\neq 0,
\end{equation}
(possibly after an up or down shifting).
The two conditions on the Jacobians (\ref{2.10}) are sufficient to allow us to
calculate $(x_{n+1}, u_{n+1})$ if $(x_{n-1}, u_{n-1}, x_n, u_n)$ are given.
Similarly, $(x_{n-1}, u_{n-1})$ can be calculated if $(x_n, u_n, x_{n+1},
u_{n+1})$ are given. The general solution of the scheme (\ref{2.9}) will hence
depend on 4 arbitrary constants and will have the form
\begin{eqnarray}
u_n & = & f(x_n, c_1, c_2, c_3, c_4)\label{2.11}\\
x_n & = & \phi(n, c_1, c_2, c_3, c_4).\label{2.12}
\end{eqnarray}

A more standard approach to difference equations would be to consider a fixed
equally spaced lattice e.g.\ with spacing $h = 1$. We can then identify the
continuous variable $x$, sampled at discrete points $x_n$, with the discrete
variable $n$:
\begin{equation}\label{2.13}
x_n = n.
\end{equation}
Instead of a difference scheme we then have a difference equation
\begin{equation}\label{2.14}
E(\{u_k\}^{n+N}_{k=n+M}) = 0,
\end{equation}
involving $K = N - M + 1$ points. Its general solution has the form
\begin{equation}\label{2.15}
u_n = f(n, c_1, c_2, \dots c_{N-M})
\end{equation}
i.e.\ it depends on $N - M$ constants.

Below, when studying point symmetries of discrete equations we will see the
advantage of considering difference systems like the system (\ref{2.2}).

\subsection{Point Symmetries of Ordinary Difference Schemes}\label{subsec2.2}

In this section we shall follow rather closely the article \cite{ref48}. We
shall define the symmetry group of an ordinary difference scheme in the same
manner as for ODEs. That is, a group of continuous local point transformations
of the form (\ref{1.2}) taking solutions of the O$\Delta$S (\ref{2.2}) into
solutions of the same scheme. The transformations considered are continuous,
and we will adopt an infinitesimal approach, as in eq.~(\ref{1.3}). We drop the
labels $i$ and $\alpha$, since we are considering the case of one independent
and one dependent variable only.

As in the case of differential equations, our basic tool will be vector fields
of the form (\ref{1.4}). In the case of O$\Delta$S they will have the form
\begin{equation}\label{2.16}
X = \xi(x, u)\partial_x + \phi(x, u)\partial_u
\end{equation}
with
\[
x \equiv x_n, \quad u \equiv u_n = u(x_n).
\]
Because we are considering point transformation, $\xi$ and $\phi$ in
(\ref{2.16}) depend on $x$ and $u$ at one point only.

The prolongation of the vector field $X$ is different than in the case
of ODEs. Instead of prolonging to derivatives, we prolong to all points of the
lattice figuring in the scheme (\ref{2.2}). Thus we put
\begin{equation}\label{2.17}
\pr X = \sum^{n+N}_{k=n+M} \xi(x_k, u_k)\partial_{x_k} + \sum^{n+N}_{k=n+M}
\phi(x_k, u_k)\partial_{u_k}.
\end{equation}
In these terms the requirement that the transformed function $\tilde u(\tilde
x)$ should satisfy the same O$\Delta$S as the original $u(x)$ is expressed by
the requirement
\begin{equation}\label{2.18}
\pr X E_a\mid_{E_1=E_2=0} = 0, \quad a = 1, 2.
\end{equation}
Since we must respect both the difference equation and the lattice, we have two
conditions (\ref{2.18}) from which to determine $\xi(x, u)$ and $\phi(x, u)$.
Since each of these functions depends on a single point  $(x, u)$ and the
prolongation (\ref{2.17}) introduces $N - M + 1$ points in space $X \times U$,
the equation (\ref{2.18}) will imply a system of determining equations for
$\xi$ and $\phi$. Moreover, in general this will be an overdetermined system of
linear functional equations that we transform into an overdetermined system of
linear differential equations \cite{ref57,ref58}.

To illustrate the method and the role of the choice of the lattice, let us
start from a simple example. The example will be that of difference equations
that approximate the ODE
\begin{equation}\label{2.19}
\stackrel{\prime\prime}{u} = 0
\end{equation}
on several different lattices.

First of all, let us find the Lie point symmetry group of the ODE (\ref{2.19}),
i.e.\ the equation of a free particle on a line. Following the algorithm of
Chapter \ref{sec1}, we put
\begin{eqnarray}
\pr^{(2)}X & = & \xi\partial_x + \phi\partial_u+
\phi^x\partial_{\stackrel{\prime}{u}} + \phi^{xx}\partial
_{\stackrel{\prime\prime}{u}}\label{2.20}\\
\phi^x & = & D_x\phi - (D_x\xi)\stackrel{\prime}{u} = \phi_x + (\phi_u-\xi_x)
\stackrel{\prime}{u}-\xi_u \smash{\stackrel{\prime}{u}}^2\nonumber\\
\phi^{xx} & = & D_x\phi^x  - (D_x\xi)\stackrel{\prime\prime}{u} = \phi_{xx} +
(2\phi_{xu} -\xi_{xx})\stackrel{\prime}{u}\nonumber\\
& & \quad + (\phi_{uu}-2\xi_{xu})\smash{\stackrel{\prime}{u}}^2 -
\xi_{uu}\smash{\stackrel{\prime}{u}}^3x + (\phi_u-2\xi_x)
\stackrel{\prime\prime}{u}\nonumber\\
& & \quad - 3\xi_u v\stackrel{\prime}{u}\stackrel{\prime\prime}{u}.\nonumber
\end{eqnarray}
The symmetry formula (\ref{1.11}) in this case reduces to
\begin{equation}\label{2.21}
\phi^{xx}\mid_{\stackrel{\prime\prime}{u}=0} = 0.
\end{equation}
Setting the coefficients of $\smash{\stackrel{\prime}{u}}^3$,
$\smash{\stackrel{\prime}{u}}^2$, $\stackrel{\prime}{u}$ and
$\smash{(\stackrel{\prime}{u}}^0)$ equal to zero, we obtain an 8 dimensional
Lie algebra, isomorphic to
$\Sl(3, \mathbb R)$ with basis
\begin{eqnarray}
X_1 & = & \partial_x, \quad X_2 = x\partial_x, \quad X_3 = u\partial_x
\label{2.22}\\
X_4 & = & \partial_u, \quad X_5 = x\partial_u, \quad X_6 = u\partial_u,
\nonumber\\
X_7 & = & x(x\partial_x+u\partial_u), \quad X_8 = u(x\partial_x+u\partial_u).
\nonumber
\end{eqnarray}

This result was of course already known to Sophus Lie. Moreover, any second
order ODE that is linear, or linearizable by a point transformation has a
symmetry algebra isomorphic to $\Sl(3, \mathbb R)$. The group $\SL(3, \mathbb
R)$ acts as the group of projective transformations of the Euclidean space
$E_2$ (with coordinates $x, u)$).

Now let us consider some difference schemes that have eq.~(\ref{2.19}) as their
continuous limit. We shall take the equation to be
\begin{equation}\label{2.23}
\frac{u_{n+1}-2u_n+u_{n-1}}{(x_{n+1}-x_n)^2} = 0.
\end{equation}
However before looking for the symmetry algebra, we multiply out the denominator
and investigate the equivalent equation
\begin{equation}\label{2.24}
E_1 = u_{n+1} - 2u_n +  u_{n-1} = 0.
\end{equation}
To this equation we must add a second equation, specifying the lattice. We
consider three different examples at first glance quite similar, but leading to
different symmetry algebras.

\begin{example}\label{exam2.1}
Free particle (\ref{2.24}) on a fixed uniform lattice. We take
\begin{equation}\label{2.25}
E_2 = x_n - hn - x_0 = 0,
\end{equation}
where $h$ and $x_0$ are fixed constants (that are not transformed by the group
(e.g.\ $h  = 1$, $x_0 = 0$).

Applying the prolonged vector field (\ref{2.17}) to eq.~(\ref{2.25}) we obtain
\begin{equation}\label{2.26}
\xi(x_n, u_n) = 0
\end{equation}
for all $x_n$ and $u_n$. Next, let us apply (\ref{2.17}) to eq.~(\ref{2.24}) and
replace $x_n$, using (\ref{2.25}) and $u_{n+1}$, using (\ref{2.24}). We obtain
\begin{eqnarray}
\lefteqn{\phi\big(h(n+1)+x_0, 2u_n-u_{n-1}\big) - 2\phi(hn+x_0, u_n)}\nonumber\\
&&\hspace{0.76in} + \phi\big(h(n-1)+x_0, u_{n-1}\big) = 0.\label{2.27}
\end{eqnarray}
Differentiating eq.~(\ref{2.27}) twice, once with respect to $u_{n-1}$, once
with respect to $u_n$, we obtain
\begin{equation}\label{2.28}
\frac{\partial^2}{\partial u_{n+1}} \phi(x_{n+1}, u_{n+1}) = 0
\end{equation}
and hence
\begin{equation}\label{2.29}
\phi(x_n,  u_n) = A(x_n)u_n + B(x_n).
\end{equation}
We substitute eq.~(\ref{2.29}) back into (\ref{2.27}) and equate coefficients
of $u_n$, $u_{n-1}$ and 1. The result is
\begin{equation}\label{2.30}
A(n+1) = A(n), \quad B(n+1) - 2B(n) + B(n-1) = 0.
\end{equation}
Hence we have
\begin{equation}\label{2.31}
A = A_0, \quad B = B_1n + B_0 = b_1x + b_0
\end{equation}
where $A_0$, $B_1$, $B_0$, $b_1$ and $b_0$ are constants. We obtain the symmetry
algebra of the O$\Delta$S (\ref{2.24}), (\ref{2.25}) and it is only
three-dimensional, spanned by
\begin{equation}\label{2.32}
X_1 = \partial_u, \quad X_2 = x\partial_u, \quad X_3 = u\partial_u.
\end{equation}
The corresponding one parameter transformation groups are obtained by
integrating these vector fields (see eq.~(\ref{1.5}), (\ref{1.6}))
\begin{eqnarray}
G_1 & : & \tilde x = x\nonumber\\
& &\tilde u(\tilde x) = u(x) + \lambda\nonumber\\
G_2 & : & \tilde x = x\label{2.33}\\
&& \tilde u(\tilde x) = u(x) + \lambda x\nonumber\\
G_3 & : & \tilde x = x\nonumber\\
&& \tilde u(\tilde x) = e^{\lambda}u(x)\nonumber
\end{eqnarray}
$G_1$ and $G_2$ just tell us that we can add an arbitrary solution of the
scheme to any given solution, $G_3$ corresponds to scale invariance of
eq.~(\ref{2.24}).
\end{example}

\begin{example}\label{exam2.2}
Free particle (\ref{2.24}) on a uniform two point lattice.

Instead of eq.~(\ref{2.25}) we define a lattice by putting
\begin{equation}\label{2.34}
E_2 = x_{n+1} - x_n = h,
\end{equation}
where $h$ is a fixed (non-transforming) constant. Note that (\ref{2.34}) tells
us the distance between any two neighbouring points but does not fix an origin
(as opposed to eq.~(\ref{2.25})).

Applying the prolonged vector field (\ref{2.17}) to eq.~(\ref{2.34}) and using
(\ref{2.34}), we obtain
\begin{equation}\label{2.35}
\xi(x_n + h, u_{n+1}) - \xi(x_n, u_n) = 0.
\end{equation}
Since $u_{n+1}$ and $u_n$ are independent, eq.~(\ref{2.35}) implies $\xi =
\xi(x)$. Moreover $\xi(x_n+h)  = \xi(x)$ so that we have
\begin{equation}\label{2.36}
\xi = \xi_0 = \const.
\end{equation}
Further, we apply $\pr X$ to eq.~(\ref{2.24}), and put $u_{n+1} = 2u_n -
u_{n-1}$, $x_{n+1} = x_n + h$, $x_{n-1} = x_n - h$ in the obtained expressions. 
As in Example~\ref{exam2.1} we find that $\phi(x, u)$ is linear in $u$ as in
(\ref{2.29}) and ultimately satisfies
\begin{equation}\label{2.37}
\phi(x, u) = au + bx + c.
\end{equation}
The symmetry algebra in this case is four-dimensional. To the basis elements
(\ref{2.32}) we add translational invariance
\begin{equation}\label{2.38}
X_4 = \partial_x.
\end{equation}
\end{example}

\begin{example}\label{exam2.3}
Free particle (\ref{2.24}) on a uniform three-point lattice.

Let us choose the lattice equation to be
\begin{equation}\label{2.39}
E_2 = x_{n+1} - 2x_n + x_{n-1} = 0.
\end{equation}
Applying $\pr X$ to $E_2$ and substituting for $x_{n+1}$ and $u_{n+1}$, we find
\begin{equation}\label{2.40}
\xi(2x_n-x_{n-1}, 2u_n-u_{n-1}) - 2\xi(x_n, u_n) + \xi(x_{n-1}, u_{n-1}) = 0.
\end{equation}
Differentiating twice with respect to $u_n$ and $u_{n-1}$, we obtain that $\xi$
is linear in $u$. Substituting $\xi = A(x) u + B(x)$ into (\ref{2.40}) we obtain
\begin{equation}\label{2.41}
\xi(x_n, u_n) = Au_n + Bx_n + C.
\end{equation}
Similarly, applying $\pr X$ to eq.~(\ref{2.24}), we obtain
\begin{equation}\label{2.42}
\phi(x_n, u_n) = Du_n + Ex_n + F.
\end{equation}
where $A$, $\dots$, $F$ are constants. Finally, we obtain a six-dimensional
symmetry algebra for the O$\Delta$S (\ref{2.24}), (\ref{2.39}) with basis
$X_1$, $\dots$, $X_6$ as in eq.~(\ref{2.22}). It has been shown \cite{ref45}
that the entire $\Sl(3, \mathbb R)$ algebra cannot be recovered on any 3 point
O$\Delta$S.
\end{example}

>From the above examples we can draw the following conclusions.
\begin{enumerate}
\item[1.] The Lie point symmetry group of an O$\Delta$S depends crucially on
both equations in the system (\ref{2.2}). In particular, if we choose a fixed
lattice, as in eq.~(\ref{2.25}) (a ``one-point lattice'') we are left with
point transformations that act on the dependent variable only.

If we wish to preserve anything like the power of symmetry analysis for
differential equations, we must either go beyond point symmetries to
generalized ones, or use lattices that are also transformed and that are
adapted to the symmetries we consider.

\item[2.] The method for calculating symmetries of O$\Delta$S  is reasonable
straightforward. It will however involve solving functional equations.
\end{enumerate}

The method can be summed up as follows
\begin{enumerate}
\item[1.] Solve equations (\ref{2.2}) for two of the quantities entering there,
to make the equations explicit. For instance, take the system (\ref{2.9}),
(\ref{2.10}). We can solve e.g.\ for $x_{n+1}$ and $u_{n+1}$ and obtain
\begin{eqnarray}
x_{n+1} & = & f_1(x_{n-1}, x_n, u_{n-1}, u_n)\label{2.43}\\
u_{n+1} & = & f_2(x_{n-1}, y_n, u_{n-1}, u_n)\nonumber
\end{eqnarray}

\item[2.] Apply the prolonged vector field (\ref{2.17}) to eq.~(\ref{2.2}) and
substitute (\ref{2.43}) for $x_{n+1}$, $u_{n+1}$. We obtain two functional
equations for $\xi$ and $\phi$ of the form
\begin{eqnarray}
\lefteqn{
\biggl\{\xi(f_1, f_2) \frac{\partial E_a}{\partial x_{n+1}} + \xi(x_n,
u_n) \frac{\partial E_a}{\partial x_n} + \xi(x_{n-1}, u_{n-1}) \frac{\partial
E_a}{\partial x_{n-1}}}\nonumber\\
&\hspace{0.5in}& + \phi(f_1, f_2)\frac{\partial E_a}{\partial u_{n+1}}
+\phi(x_u, u_n) \frac{\partial E_a}{\partial u_n}\label{2.44}\\
&\hspace{0.5in}& + \phi(x_{n-1}, u_{n-1})
\frac{\partial E_a}{\partial u_{n-1}}\biggr\}\biggm|\nonumber_{\stackrel
{x_{n+1}=f_1}{u_{n+1}=f_2}} = 0 \quad a = 1,2\nonumber
\end{eqnarray}

\item[3.] Assume that the functions $\xi$, $\phi$, $E_1$ and $E_2$ are
sufficiently smooth and differentiate eq.~(\ref{2.44}) with respect to the
variables $x_k$ and $u_k$ so as to obtain differential equations for $\xi$ and
$\phi$. If the original equations are polynomial in all quantities we can thus
obtain single term differential equations form (\ref{2.44}). These we must
solve, then substitute back into (\ref{2.44}) and solve this equation.
\end{enumerate}

We will illustrate the procedure on several examples in Section~\ref
{subsec2.3}.

\subsection{Examples of Symmetry Algebras of O$\Delta$S}\label{subsec2.3}

\begin{example}\label{exam2.4}
Monomial nonlinearity on a uniform lattice.

Let us first consider the nonlinear ODE
\begin{equation}\label{2.45}
\stackrel{\prime\prime}{u} - u^N = 0, \quad N \neq 0, 1.
\end{equation}
For $N \neq - 3$ eq.~(\ref{2.45}) is invariant under a two-dimensional Lie
group, the Lie algebra of which is given by
\begin{equation}\label{2.46}
X_1 = \partial_x, \quad X_2 = (N-1)x\partial_x - 2u\partial_u
\end{equation}
(translations and dilations).
For $N = -3$ the symmetry algebra is three-dimensional, isomorphic to $\Sl(3,
\mathbb R)$, i.e.\ it contains a third element, additional to (\ref{2.46}). A
convenient basis for the symmetry algebra of the equation
\begin{equation}\label{2.47}
\stackrel{\prime\prime}{u} - u^{-3} = 0
\end{equation}
is
\begin{equation}\label{2.48}
X_1 = \partial_x, \quad X_2 = 2x\partial_x + u\partial_u, \quad X_3 = x(x
\partial_x+u\partial_u).
\end{equation}

A very natural O$\Delta$S that has (\ref{2.45}) as its continuous limit is
\begin{eqnarray}
E_1 & = & \frac{u_{n+1}-2u_n+u_{n-1}}{(x_{n+1}-x_n)^2} - u^N_n =  0 \quad N
\neq 0, 1\label{2.49a}\\
E_2 & = & x_{n+1} - 2x_n +  x_{n-1} = 0\label{2.49b}
\end{eqnarray}

Let us now apply the symmetry algorithm described in Chapter \ref{subsec2.2} to
the system (\ref{2.49a}) and (\ref{2.49b}). To illustrate the method, we shall
present all calculations in detail.

First of all, we choose two variables that will be substituted in
eq.~(\ref{2.18}), once the prolonged vector field (\ref{2.17}) is applied to
the system (\ref{2.49a}) and (\ref{2.49b}), namely
\begin{eqnarray}
x_{n+1} & = & 2x_n - x_{n-1}\label{2.50}\\
u_{n+1} & = & (x_n-x_{n-1})^2u^N_n + 2u_n - u_{n-1}\nonumber
\end{eqnarray}
We apply $\pr X$ of (\ref{2.17}) to eq.~(\ref{2.49b}) and obtain
\begin{equation}\label{2.51}
\xi(x_{n+1}, u_{n+1}) - 2\xi(x_n, u_n) + \xi(x_{n-1}, u_{n-1}) = 0
\end{equation}
where, $x_n$, $u_n$ $x_{n-1}$, $u_{n-1}$ are independent, but $x_{n+1}$,
$u_{n+1}$ are expressed in terms of these quantities, as in eq.~(\ref{2.50}).
Taking this into acccount, we differentiate (\ref{2.51}) first with respect to
$u_{n-1}$, then $u_n$. We obtain successively
\begin{eqnarray}
- \xi_{,u_{n+1}}(x_{n+1}, u_{n+1}) + \xi_{,u_{n-1}}(x_{n-1}, u_{n-1}) & = &0
\label{2.52}\\
\{N(x_n-x_{n-1})^2 u^{N-1}_n +2\} \xi_{,u_{n+1}u_{n+1}} (x_{n+1}, u_{n+1}) 
& = & 0.\label{2.53}
\end{eqnarray}
Eq.~(\ref{2.53}) is the desired one-term equation. It implies
\begin{equation}\label{2.54}
\xi(x, u) = a(x)u + b(x)
\end{equation}
Substituting (\ref{2.54}) into (\ref{2.52}) we obtain
\begin{equation}\label{2.55}
- a(2x_n-x_{n-1}) + a(x_{n-1}) = 0.
\end{equation}
Differentiating with respect to $x_n$, we obtain $a = a_0 = \const$. Finally,
we substitute (\ref{2.54}) with $a = a_0$ into (\ref{2.51}) and obtain
\begin{equation}\label{2.56}
a = 0, \quad b(2x_n-x_{n-1}) - 2b(x_n) + b(x_{n-1}) = 0
\end{equation}
and hence
\begin{equation}\label{2.57}
\xi = b = b_1x + b_0
\end{equation}
where $b_0$ and $b_1$ are constants. To obtain the function $\phi(x_n, u_n)$,
we apply $\pr  X$ to eq.~(\ref{2.49a}) and obtain
\begin{eqnarray}
\lefteqn{\phi(x_{n+1}, u_{n+1}) - 2\phi(x_n, u_n) + \phi(x_{n-1}, u_{n-1})}
\nonumber\\
&&\hspace{1in}-(x_n-x_{n-1})^2 [N\phi(x_n, u_n)u^{N-1}_n + 2b_1u^N_n] =
0.\label{2.58}
\end{eqnarray}
Differentiating successively with respect to $u_{n-1}$ and $u_n$ (taking
(\ref{2.50}) into account) we obtain
\begin{eqnarray}
- \phi_{,u_{n+1}}(x_{n+1}, u_{n+1}) + \phi_{,u_{n-1}}(x_{n-1}, u_{n-1}) & = &
0\label{2.59}\\
\{N(x_n-x_{n-1})^2u^N_n +2\} \phi_{,u_{n+1}u_{n+1}} & = & 0\label{2.60}
\end{eqnarray}
and hence
\begin{equation}\label{2.61}
\phi = \phi_1u + \phi_0(x), \quad \phi_1 = \const.
\end{equation}
Eq.~(\ref{2.58}) now reduces to
\begin{eqnarray}
\lefteqn{\phi_0(2x_n-x_{n-1}) - 2\phi_0(x_n) + \phi_0(x_{n-1})}\nonumber\\
&&\hspace{0.50in} -(x_n-x_{n-1})^2 \{(N-1)\phi_1 + 2b_1\}u^N_n\nonumber\\
&&\hspace{0.50in} - N(x_n-x_{n-1})^2\phi_0u^{N-1}_n = 0.\label{2.62}
\end{eqnarray}
We have $N \neq 0, 1$ and hence (\ref{2.62}) implies
\begin{equation}\label{2.63}
\phi_0 = 0, \quad (N-1)\phi_1 + 2b_1 = 0.
\end{equation}
We have thus proven that the symmetry algebra of the O$\Delta$S (\ref{2.49a})
and (\ref{2.49b}) is the same as that of the ODE (\ref{2.45}), namely the
algebra (\ref{2.46}).

We mention that the value $N = -3$ is not distinguished here and the system
(\ref{2.49a}) and (\ref{2.49b}) is not invariant under $\SL(3, \mathbb R)$ for
$N = -3$. Actually, a difference scheme invariant under $\SL(3, \mathbb R)$ does
exist and it will have eq.~(\ref{2.47}) as its continuous limit. It will
however not have the form (\ref{2.48}) and the lattice will not be uniform
\cite{ref45,ref46}.

Had we taken a two-point lattice, $x_{n+1}-x_n = h$ with $h$ fixed, instead of
$E_2 = 0$ as in (\ref{2.49b}), we would only have obtained translational
invariance for the equation (\ref{2.49a}) and lost the dilational invariance
represented by $X_2$ of eq.~(\ref{2.46}).
\end{example}

\begin{example}\label{exam2.5}
A nonlinear O$\Delta$S on a uniform lattice
\begin{eqnarray}
E_1 & = & \frac{u_{n+1}-2u_n+u_{n-1}}{(x_{n+1}-x_n)^2} - f\left(\frac{u_n-
u_{n-1}}{x_n-x_{n-1}}\right) = 0,\label{2.64a}\\
E_2 & = & x_{n+1} -2x_n + x_{n-1} = 0,\label{2.64b}
\end{eqnarray}
where $f(z)$ is some sufficiently smooth function satisfying
\begin{equation}\label{2.65}
\stackrel{\prime\prime}{f}(z) \neq 0.
\end{equation}
The continuous limit of eq.~(\ref{2.64a}) and (\ref{2.64b}) is
\begin{equation}\label{2.66}
\stackrel{\prime\prime}{u} - f(\stackrel{\prime}{u}) = 0
\end{equation}
and is invariant under a two-dimensional group with Lie algebra
\begin{equation}\label{2.67}
X_1 = \partial_x, \quad X_2 = \partial_u
\end{equation}
for any function $f(\stackrel{\prime}{u})$. For certain functions $f$ the
symmetry group is three-dimensional, where the additional basis element of the
Lie algebra is
\begin{equation}\label{2.68}
X_3 = (ax+bu)\partial_x + (cx+du)\partial_u.
\end{equation}
The matrix
\begin{equation}\label{2.69}
M = \left(\begin{array}{ll} 
a\quad  b\\ c \quad d\end{array}\right)
\end{equation}
can be transformed to Jordan canonical form and a different function $f(z)$ is
obtained for each canonical form.

Now let  us consider the discrete system (\ref{2.64a}) and (\ref{2.64b}).
Before applying $\pr X$ to this system we choose two variables to substitute in
eq.~(\ref{2.18}), namely
\begin{eqnarray}
x_{n+1} & = & 2x_n - x_{n-1}\label{2.70}\\
u_{n+1} & = & 2u_n - u_{n-1} + (x_n-x_{n-1})^2 f\left(\frac{u_n-u_{n-1}}
{x_n-x_{n-1}}\right).\nonumber
\end{eqnarray}
Applying $\pr X$ to eq.~(\ref{2.64b}) we obtain eq.~(\ref{2.51}) with $x_{n+1}$
and $u_{n+1}$ as in eq.~(\ref{2.70}). Differentiating twice, with respect to
$u_{n-1}$ and $u_n$ respectively, we obtain
\begin{equation}\label{2.71}
\xi_{, u_{n+1}u_{n+1}}[1+(x_n-x_{n-1})f'] [2+(x_n-x_{n-1})f'] + \xi_{, u_{n+1}}
f'' = 0.
\end{equation}
For $f'' \neq 0$ the only solution is $\xi_{, u_{n+1}} = 0$, i.e. $\xi =
\xi(x)$. Substituting back into (\ref{2.51}), we obtain
\begin{equation}\label{2.72}
\xi = \alpha x + \beta
\end{equation}
with $\alpha = \const$, $\beta = \const$.

Now let us apply $\pr X$ to $E_1$ of eq.~(\ref{2.64a}) and (\ref{2.64b}) and
replace $x_{n+1}$,
$u_{n+1}$ as in eq~(\ref{2.70}). We obtain the equation
\begin{eqnarray}
\lefteqn{\phi(x_{n+1}, u_{n+1}) - 2\phi(x_n, u_n) + \phi(x_{n-1},
u_{n-1}) = 2\alpha(x_n-x_{n-1})^2f(z)}\nonumber\\ 
&& \hspace{.65in}+
(x_n-x_{n-1})^2f'(z)\left(\frac{\phi(x_n, u_n)-\phi(x_{n-1},
u_{n-1})}{x_n-x_{n-1}} -\alpha z\right)\label{2.73}
\end{eqnarray}
with $\alpha$ as in eq.~(\ref{2.72}). Thus, we only need to distinguish between
$\alpha = 0$ and $\alpha = 1$. Eq.~(\ref{2.73}) is a functional equation,
involving two unknown functions $\phi$ and $f$. There are only four independent
variables involved, $x_n, x_{n-1}, u_n$ and $u_{n-1}$. We simplify (\ref{2.73})
by introducing new variables $\{x, u, h, z\}$, putting
\begin{eqnarray}
x_n & = & x, \quad x_{n+1} = x + h, \quad x_{n-1} = x - h\label{2,74}\\
u_n & = & u,\quad u_{n-1} = u - hz, \quad u_{n+1} = u + hz + h^2f(z),\nonumber
\end{eqnarray}
where we have used eq.~(\ref{2.70}) and defined
\begin{equation}
z = \frac{u_n-u_{n-1}}{x_n-x_{n-1}} \quad h = x_{n+1} - x_n.
\end{equation}
Eq.~(\ref{2.73}) in these variables is
\begin{eqnarray}
\lefteqn{\phi\big(x+h, u+hz+h^2f(z)\big) - 2\phi(x, u) + \phi(x-h,
u-hz)}\nonumber\\
&&\hspace{.25in}=2\alpha h^2f(z)
+ h^2f'(z)\biggl[\frac{\phi(x, u)-\phi(x-h, u-hz)}{h}-\alpha
z\biggr].\label{2.76}
\end{eqnarray}

First of all, we notice that for any function $f(z)$ we have two obvious
symmetry elements, namely $X_1$ and $X_2$ of eq.~(\ref{2.67}), corresponding to
$\alpha = 0$, $\beta = 1$ in (\ref{2.76}) (and (\ref{2.72})) and $\phi = 0$ and
$\phi = 1$, respectively. Eq.~(\ref{2.76}) is quite difficult to solve
directly. However, any three-dimensional Lie algebra of vector fields in 2
variables, containing $\{X_1, X_2\}$ of eq.~(\ref{2.67}) as a subalgebra, must
have $X_3$ of eq.~(\ref{2.68}) as its third element. Moreover, eq.~(\ref{2.72})
shows that we have $b = 0$ in eq.~(\ref{2.68}) and (\ref{2.69}). In
(\ref{2.76}) we put $\alpha = a$ and
\begin{equation}\label{2.77}
\phi(x, u) = cx + du.
\end{equation}
Substituting into eq.~(\ref{2.76}) we obtain
\begin{equation}\label{2.78}
(d-2a)f(z)  = [c+(d-a) z]f'(z).
\end{equation}
>From eq.~(\ref{2.78}) we obtain two types of solutions:

\noindent For $d \neq a$ we have
\begin{equation}\label{2.79}
f = f_0[(d-a)z+c]^{(d-2a)/(d-a)}, \quad c \neq 0.
\end{equation}

\noindent For $d = a$, we have
\begin{equation}\label{2.80}
f = f_0e^{-(a/c) x}.
\end{equation}
With no loss of generality we could have taken the matrix (\ref{2.69}) with $b
= 0$ to Jordan cannonical form and we would have obtained two different cases,
simplifying (\ref{2.79}) and (\ref{2.80}), respectively. They are
\begin{eqnarray}
f & = & f_0z^N, \quad X_3 = x\partial_x + \frac{N-2}{N-1} u\partial u, \quad
N \neq 1\label{2.81}\\
f & = & f_0 e^{-z}, \quad X_3 = x\partial_x + (x+u)\partial u.\label{2.82}
\end{eqnarray}
The result can be stated as follows. The O$\Delta$S (\ref{2.64a}) and
(\ref{2.64b}) is always invariant under the group generated by $\{X_1, X_2\}$
as in (\ref{2.67}). It is invariant under a three-dimensional group with
algebra including $X_3$ as in eq.~(\ref{2.68}) if $f$ satisfies
eq.~(\ref{2.78}), i.e.\ has the form (\ref{2.81}), or (\ref{2.82}). These two
cases also exist in the continuous limit. However, one more case exists in the
continuous limit, namely
\begin{equation}\label{2.83}
\stackrel{\prime\prime}{u} = \big(1+(\stackrel{\prime}{u})^2\big)^{3/2}e^
{k\arctan \, \stackrel{\prime}{u}}
\end{equation}
with
\begin{equation}\label{2.84}
X_3 = (kx+u)\partial_x + (ku-x)\partial u.
\end{equation}
This equation can also be discretized in a symmetry preserving way
\cite{ref45}, not however on the uniform lattice (\ref{2.64b}).
\end{example}

\section{Lie Point Symmetries of Partial Difference Schemes}\label{sec3}

\subsection{Partial Difference Schemes}\label{subsec3.1}

In this chapter we generalize the results of Chapter~\ref{sec2} to the case of
two discretely varying independent variables. We follow the ideas and notation
of Ref.~\cite{ref49}. The generalization to $n$ variables is
immediate, though cumbersome. Thus, we will consider a Partial Difference Scheme
(P$\Delta$S), involving one continuous function of two continuous variables
$u(x, t)$. The variables $(x, t)$ are sampled on a two-dimensional lattice,
itself defined by a system of compatible relations between points.
Thus, a lattice will be an a priori infinite set of points $P_i$ lying in the
real plane $\mathbb R^2$. The points will be labelled by two discrete
subscripts $P_{m, n}$ with $-\infty < m < \infty$, $- \infty < n < \infty$. The
cartesian coordinates of the point $P_{mn}$ will be denoted $(x_{mn},
t_{mn})$ [or similarly any other coordinates $(\alpha_{mn}, \beta_{mn})$].

\begin{figure}
\begin{center}
\includegraphics{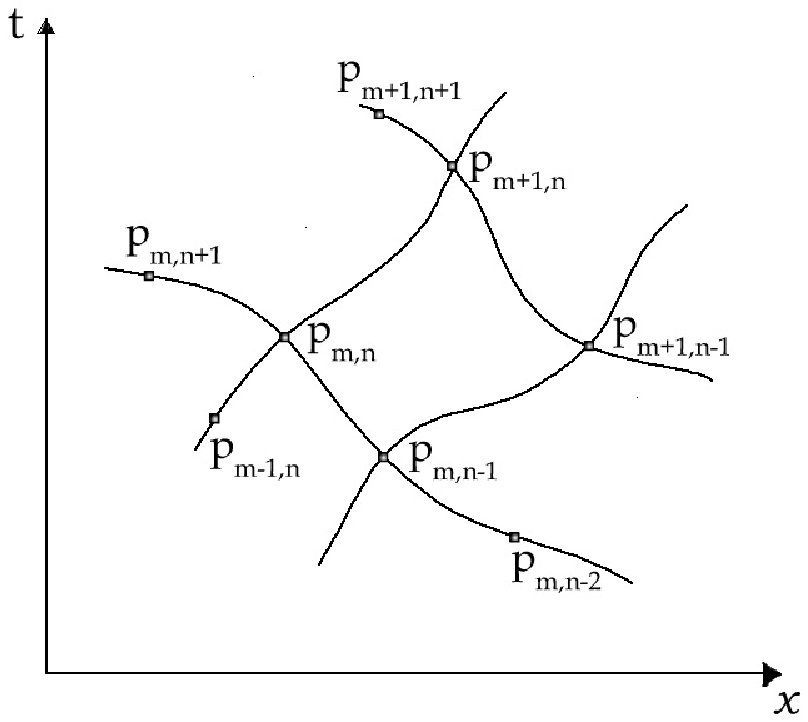}
\end{center}
\caption{}
\label{fig1}
\end{figure}

A two-variable P$\Delta$S will be a set of five relations between the quantities
$\{x, t, u\}$ at a finite number of points. We choose a reference point $P_{mn}
\equiv P$ and two families of curves intersecting at the points of the lattice.
The labels $m = m_0$ and $n = n_0$ will parametrize these curves (see
Fig~\ref{fig1}). To define an orientation of the curves, we specify 
\begin{equation}\label{3.1}
x_{m+1n} - x_{m, n} \equiv h_m > 0, \quad t_{mn+1}-t_{mn} \equiv h_n > 0
\end{equation}
at the original reference point.

The actual curves and the entire P$\Delta$S are specified by the 5 relations
\begin{equation}\label{3.2}
\begin{array}{c}
E_a(\{x_{m+i, n+j}, t_{m+i, n+j}, u_{m+i, n+j}\}) = 0\\[\jot]
1 \leq a \leq 5 \quad i_1 \leq i \leq i_2 \quad j_i \leq j \leq j_2.
\end{array}
\end{equation}

In the continuous limit, if one exists, all five equations (\ref{3.2}) are
supposed to reduce to a single PDE, e.g. $E_1 = 0$ can reduce to the PDE and
$E_a=0$, $a \geq 2$ to $0 = 0$. The orthogonal uniform lattice of
Fig.~\ref{fig2} is clearly a special case of that on Fig.~\ref{fig1}.

\begin{figure}
\begin{center}
\setlength{\unitlength}{1.5pt}
\begin{picture}(130,110)(-12.5,-3)
\put(0,10){\line(1,0){110}}
\put(10,0){\line(0,1){100}}
\put(10,40){\line(1,0){80}}
\put(10,80){\line(1,0){90}}
\multiput(30,7.5)(0,5){15}{\line(0,1){2.5}}
\put(60,7.5){\line(0,1){72.5}}
\multiput(90,7.5)(0,5){15}{\line(0,1){2.5}}
\put(30,40){\circle{5}}
\put(90,40){\circle{5}}
\put(60,80){\circle{5}}
\put(60,40){\circle*{5}}
\put(5,40){\makebox(0,0){$n$}}
\put(-5,80){\makebox(0,0){$n+1$}}
\put(30,0){\makebox(0,0){$m-1$}}
\put(60,0){\makebox(0,0){$m$}}
\put(90,0){\makebox(0,0){$m+1$}}
\put(5,105){\makebox(0,0){$t$}}
\put(115,5){\makebox(0,0){$x$}}
\end{picture}
\end{center}
\caption{}
\label{fig2}
\end{figure}

Some independence conditions must be imposed on the system (\ref{3.2}) e.g.
\begin{equation}\label{3.3}
|J| = \left|\frac{\partial(E_1, \dots, E_5)}{\partial(x_{m+i_2, n}, t_{m+i_2
n}, x_{m, n+j_2}, t_{m, n+j_2},u_{m+i_2, n+j_2})}\right| \neq 0.
\end{equation}
This condition allows us to move upward and to the right along the curves
passing through $P_{m, n}$. Moreover, compatibility of the five equations
(\ref{3.2}) must be assured.

As an example of a P$\Delta$S let us consider the linear heat equation
on a uniform and orthogonal lattice. The heat equation in the continuous case is
\begin{equation}\label{3.4}
u_t = u_{xx}.
\end{equation}

An approximation on a uniform orthogonal lattice is given by the five
equations
\begin{eqnarray}
E_1  &=& \frac{u_{mn+1}-u_{mn}}{h_2} - \frac{u_{m+1n}-2u_{mn}+u_{m-1n}}
{(h_1)^2} = 0\label{3.5}\\
E_2  &=&  x_{m+1, n}-x_{m, n}-h_1 = 0 \quad\quad E_3 = t_{m+1, n} -t_{m, n}  =
0\label{3.6}\\
E_4 &=& x_{m, n+1} -x_{mn} = 0 \quad\quad E_5 = t_{m, n+1} - t_{m, n} - h_2 =
0\label{3.7}.
\end{eqnarray}
Equations (\ref{3.6}) can of course be integrated to give the standard
expressions
\begin{equation}\label{3.8}
x_{mn} = h_1m + x_0 \quad t_{mn} = h_2n + t_0.
\end{equation}
Notice that $h_1$ and $h_2$ are constants that cannot be scaled (they are fixed
in eq.~(\ref{3.6}). On the other hand $(x_0, t_0)$ are integration constants
and are thus not fixed by the system (\ref{3.6}), (\ref{3.7}). As written, these
equations are invariant under translations, but not under dilations.

Finally, we remark that the usual fixed lattice condition is obtained from
(\ref{3.7}) by putting $x_0 = t_0 = 0$, $h_1 = h_2 = 1$ and identifying
\begin{equation}\label{3.9}
x = m, \quad t = n.
\end{equation}

Though the above example is essentially trivial, it brings out several points.
\begin{enumerate}
\item[1.] Four equations are indeed needed to specify a two-dimensional lattice
and to allow us to move along the coordinate lines.

\item[2.] In order to solve the P$\Delta$S (\ref{3.5}), (\ref{3.6}) for $h_1$
and $h_2$ given, we must specify for instance $\{x_{mn}, t_{mn}, u_{mn}, u_{m+1,
n}, u_{m-1n}\}$. Then we can directly calculate $\{x_{m+1, n}, t_{m+1n}\}$,
$\{x_{mn+1}, t_{m, n+1}\}$ from eq.~(\ref{3.6}). In order to calculate the
coordinates of the fourth point figuring in eq.~(\ref{3.5}), namely
$\{x_{mn-1}, t_{m,n-1}\}$ we must shift eq.~(\ref{3.6}) down by one unit in $m$.

\item[3.] The Jacobian condition (\ref{3.3}) allowing us to perform these
calculations, is obviously satisfied, since we have
\begin{equation}\label{3.10}
\left|\frac{\partial(E_1, E_2, E_3, E_4, E_5)}{\partial(x_{m+1, n}, t_{m+1, n},
x_{m, n+1}, t_{mn+1}, u_{mn+1})}\right| = 1.
\end{equation}
\end{enumerate}

A partial difference scheme with one dependent and $n$ independent variables
will involve $n^2+1$ relations between the variables $(x_1, x_2, \dots x_n,
u)$, evaluated at a finite number of points.

\subsection{Symmetries of Partial Difference Schemes}\label{subsec3.2}

As in the case of O$\Delta$S treated in Chapter~\ref{sec2}, we shall restrict
ourselves to point transformations
\begin{equation}\label{3.11}
\tilde x = F_{\lambda}(x, t, u) \quad \tilde t = G_{\lambda}(x, t, u), \quad
\tilde u = H_{\lambda}(x, t, u).
\end{equation}
The requirement is that $\tilde u_{\lambda}(\tilde x, \tilde t)$ should be a
solution, whenever it is defined and whenever $u(x, t)$ is a solution. The
group action (\ref{3.11}) should be defined and invertible, at least locally,
in some neighbourhood of the reference point $P_{mn}$, including all points
$P_{m+i, n+j}$ involved in the system (\ref{3.2}).

As in the case of a single independent variable we shall consider infinitesimal
transformations that allow us to use Lie algebraic techniques. Instead of
transformations (\ref{3.11}) we consider
\begin{eqnarray}
\tilde x &=& x + \lambda\xi(x, t, u),\nonumber\\ 
\tilde t &=& t + \lambda\tau(x, t, u),\label{3.12}\\
\tilde u &=& u + \lambda\phi(x, t, u) \quad|\lambda| << 1.\nonumber
\end{eqnarray}

Once the functions $\xi$, $\tau$ and $\phi$ are determined from the invariance
requirement, then the actual transformations (\ref{3.11}) are determined by
integration, as in eq.~(\ref{1.5}), (\ref{1.6}).

The transformations act on the entire space $(x, t, u)$, at least locally. This
means that the same function $F$, $G$ and $H$ in eq.~(\ref{3.11}), or $\xi$,
$\tau$ and $\phi$ in eq.~(\ref{3.12}) determine the transformations of all
points.

We formulate the problem of determining the symmetries (\ref{3.12}), and
ultimately (\ref{3.11}), in terms of a Lie algebra of vector fields of the form
\begin{equation}\label{3.13}
\widehat X = \xi(x, t, u)\partial x + \tau(x, t, u)\partial_t + \phi(x, t,
u)\partial_u,
\end{equation}
where $\xi$, $\tau$ and $\phi$ are the same as in eq.~(\ref{3.12}). The
operator (\ref{3.13}) acts at one point only, namely $(x, t, u) \equiv (x_{mn},
t_{mn}, u_{mn})$. Its prolongation will act at all points figuring in the
system(\ref{3.2}) and we put
\begin{eqnarray}
\lefteqn{\pr X = \sum_{j, k}[\xi(x_{jk}, t_{jk}, u_{jk})\partial_{x_{jk}} +
\tau(x_ {jk}, t_{jk}, u_{jk})\partial_{t_{jk}}}\label{3.14}\\
&\hspace{2in}&+ \phi(x_{jk}, t_{jk}, u_{jk})\partial_{u_{jk}}],\nonumber
\end{eqnarray}
where the sum is over all points figuring in eq.~(\ref{3.2}). To simplify
notation we put
\begin{eqnarray}
\xi_{jk} &\equiv& \xi(x_{jk}, t_{jk}, u_{jk}), \quad \tau_{jk} \equiv \tau
(x_{jk}, t_{jk}, u_{jk})\label{3.15}\\
\phi_{jk} &\equiv& \phi(x_{jk}, t_{jk}, u_{jk}).\nonumber
\end{eqnarray}

The functions $\xi$, $\tau$, and $\phi$ figuring in eq.~(\ref{3.13}) and
(\ref{3.14}) are determined from the invariance condition
\begin{equation}\label{3.16}
\pr\widehat X E_a\mid_{E_1=\dots=E_5=0} = 0 \quad a = 1, \dots 5.
\end{equation}

It is eq.~(\ref{3.16}) that provides an algorithm for determining the symmetry
algebra, i.e.\ the coefficients $\xi$, $\tau$ and $\phi$.

The procedure is the same as in the case of ordinary difference schemes,
described in Chapter~\ref{sec2}. In the case of the system (\ref{3.2}), we
proceed as follows:
\begin{enumerate}
\item[1.] Choose 5 variables $v_a$ to eliminate from the condition (\ref{3.16})
and express them in terms of the other variables, using the system (\ref{3.2})
and the Jacobian condition (\ref{3.3}). For instance, we can choose
\begin{eqnarray}
v_1 & = & x_{m+i_2, n}, \quad v_2 = t_{m+i_2, n}\label{3.17}\\
v_3 & = & x_{m, n+j_2}, \quad v_4 = t_{m, n+j_2}, \quad v_5 = u_{m+i_2,
j+i_2}\nonumber
\end{eqnarray}
and use (\ref{3.2}) to express
\begin{equation}
\begin{array}{c}
v_a  =  v_a(x_{m+i, n+j}, t_{m+i, n+j}, u_{m+i, n+j})\label{3.18}\\[\jot]
i_1  \leq  i \leq i_2 - 1, \quad j_1 \leq j \leq j_2 - 1.\nonumber
\end{array}
\end{equation}
The quanties $v_a$ must be chosen consistently. None of them can be a shifted
value of another one (in the same direction). No relations between the
quantities $v_a$ should follow from the system (\ref{3.2}). Once eliminated
from eq.~(\ref{3.15}), they should not reappear due to shifts. For instance,
the choice (\ref{3.17}) is consistent if $m + i_2$ and $n + j_2$ are the highest
values of these labels that figure in eq.~(\ref{3.2}).

\item[2.] Once the quantities $v_a$ are eliminated from the system
(\ref{3.16}), using (\ref{3.18}), each remaining value of $x_{i, k}$, $t_{i,
k}$ and $u_{i, k}$ is independent. Each of them can figure in the corresponding
functions $\xi_{ik}$, $\tau_{ik}$, $\phi_{ik}$ (see eq.~(\ref{3.15})), in
the functions $E_a$ directly, or via the expressions $v_a$, in the functions
$\xi$, $\tau$ and $\phi$ with different labels. This provides a system of five
functional equations for $\xi$, $\tau$ and $\phi$.

\item[3.] Assume that the dependence of $\xi$, $\tau$ and $\phi$ on $x$, $t$
and $u$ is analytic. Convert the obtained functional equations into
differential equations by differentiating with respect to $x_{ik}$, $t_{ik}$, or
$u_{ik}$. This provides an overdetermined system of differential equations that
we must solve. If possible, use multiple differentiations to obtain single term
differential equations that are easy to solve.

\item[4.] Substitute the solution of the differential equations back into the
original functional equations and solve these. The differential equations are
consequences of the functional ones and will hence have more solutions. The
functional equations will provide further restrictions on the constants and
arbitrary functions obtained when integrating the differential consequences.

Let us now consider examples on different lattices. 
\end{enumerate}

\subsection{The Discrete Heat Equation}\label{subsec3.3}

\subsubsection{The Continuous Heat Equation}\label{subsubsec3.3.1}

The symmetry group of the continuous heat equation (\ref{3.4}) is well
known \cite{ref1}. Its symmetry algebra has the structure of a semidirect sum
\begin{equation}\label{3.19}
L = L_0 \semisum L_1,
\end{equation}
where $L_0$ is six-dimensional and $L_1$ is an infinite dimensional ideal
representing the linear superposition principle (present for any linear PDE). A
convenient basis for this algebra is given by the vector fields
\begin{eqnarray}
P_0 &=& \partial_t, \quad D = 4t\partial_t + 2x\partial_x
+ u\partial_u\nonumber\\
K &=& 4t(t\partial_t+x\partial_x) + (x^2+2t)u\partial_u\label{3.20}\\
P_1 &=& \partial_x, B = 2t\partial_x + xu\partial_u, \quad
W = u\partial_u\nonumber\\
S &=& S(x, t)\partial_u, \quad S_t - S_{xx} = 0.\label{3.21}
\end{eqnarray}
The $\Sl(2, \mathbb R)$ subalgebra $\{P_0, D, K\}$ represents time
translations, dilations and ``expansions''. The Heisenberg subalgebra $\{P_1,
B, W\}$ represents space translations, Galilei boosts and the possibility of
multiplying a solution $u$ by a constant. The presence of $\widehat S$ simply
tells us that we can add a solution to any given solution. Thus, $\widehat S$
and $\widehat W$ reflect linearity, $\widehat P_0$ and  $\widehat P_1$ the fact
that the equation is autonomous (constant coefficients).

\subsubsection{Discrete Heat Equation on Fixed Rectangular Lattice}
\label{subsubsec3.3.2}

Let us consider the discrete heat equation (\ref{3.5}) on the four point
uniform orthogonal lattice (\ref{3.6}), (\ref{3.7}). We apply the prolonged
operator (\ref{3.4}) to the equations for the lattice and obtain
\begin{equation}\label{3.22}
\begin{array}{c}
\xi(x_{m+1n}, t_{m+1n}, u_{m+1n}) - \xi(x_{mn}, t_{mn}, u_{mn}) = 0\\[\jot]
\xi(x_{m, n+1}, t_{m, n+1}, u_{m, n+1}) - \xi(x_{mn}, t_{mn}, u_{mn}) = 0               
\end{array}
\end{equation}
and similarly for $\tau(x, t, u)$. The quantities $v_i$ of eq.~(\ref{3.17}) can
be chosen to be
\begin{equation}\label{3.23}
\begin{array}{c}
v_1 = x_{m+1, n} \quad v_2 = t_{m+1, n} \quad v_3 = x_{m, n+1},\\[\jot]
v_4 = t _{m, n+1}, \quad v_5 = u_{m, n+1}.
\end{array}
\end{equation}
However, in (\ref{3.22}) $u_{m+1, n}$ and $u_{m, n+1}$ cannot be expressed in
terms of $u_{mn}$, since eq.~(\ref{3.5}) also involves $u_{m-1, n}$.
Differentiating (\ref{3.22}) with respect to e.g. $u_{mn}$. we find that $\xi$
cannot depend on $u$:
\begin{equation}\label{3.24}
\frac{\partial\xi(x_{mn}, t_{mn}, u_{mn})}{\partial u_{mn}} = 0.
\end{equation}
Since we have $t_{n+1n} = t_{mn}$ and $x_{mn+1} = x_{mn}$ the two equations
(\ref{3.22}) yield
\begin{equation}\label{3.25}
\frac{\partial\xi_{mn}}{\partial x_{mn}} = 0, \quad
\frac{\partial\xi_{mn}}{\partial t_{mn}} = 0,
\end{equation}
respectively. The same is obtained for the coefficient $\tau$, so finally we
have
\begin{equation}\label{3.26}
\xi = \xi_0, \quad \tau = \tau_0,
\end{equation}
where $\xi_0$ and $\tau_0$ are constants.

Now let us apply $\pr X$ to eq.~(\ref{3.5}). We obtain
\begin{equation}\label{3.27}
\phi_{mn+1} - \phi_{mn} - \frac{h_2}{(h_1)^2} (\phi_{m+1n}-2\phi_{mn} +
\phi_{m-1, n}) = 0.
\end{equation}
In more detail, eliminating the quantities $v_a$ in eq~(\ref{3.23}) we have
\begin{eqnarray}
\lefteqn{\phi\left(x_{mn}, t_{mn}+h_2, u_{m, n}+\frac{h_2}{h^2_1}(u_{m+1,
n}-2u_{m, n} +u_{m-1, n})\right)}\nonumber\\
&&-\phi(x_{m, n},t_{m, n},u_{m, n})
-\frac{h_2}{h^2_1}[\phi(x_{mn}+h_1, t_{mn}, u_{m+1, n})
-2\phi(x_{m, n}, t_{m, n}, u_{m, n})\nonumber\\ 
&&\hspace{2in}+ \phi(x_{mn}-h_1, t_{mn}, u_{m-1, n})] = 0.\label{3.28}
\end{eqnarray}
We differentiate eq.~(\ref{3.28}) twice, with respect to $u_{m+1, n}$ and
$u_{m-1, n}$ respectively. We obtain
\begin{equation}
\frac{\partial^2\phi_{mn+1}}{\partial u_{m, n+1}^2} = 0,
\end{equation}
that is
\begin{equation}\label{3.30}
\phi_{mn} = A(x_{mn}, t_{m, n}) u_{mn} + B(x_{m, n}, t_{m, n}).
\end{equation}
We substitute $\phi_{mn}$ of eq.~(\ref{3.30}) back into eq.~(\ref{3.28}) and
set the coefficients of $u_{m+1, n}$, $u_{mn}$, $u_{m-1n}$ and 1 equal to zero
separately. From the resulting determining equations we find that $A(x_{mn},
t_{mn}) = A_0$ must be constant and that $B(x, t)$ must satisfy the 
discrete heat equation (\ref{3.5}). The result is that the symmetry algebra of
the system (\ref{3.5}) - (\ref{3.7}) is very restricted. It is generated by
\begin{equation}\label{3.31}
P_0 = \partial_t, \quad P_1 = \partial_x, \quad W = u\partial_u, \quad
S = S(x, t)\partial _u
\end{equation}
and reflects only the linearity of the system and the fact that it is
autonomous.

The dilations, expansions and Galilei boosts, generated by $D$,
$K$ and $B$ of eq.~(\ref{3.20}) in the continuous case are
absent on the lattice (\ref{3.6}) and (\ref{3.7}). Other lattices will allow
other symmetries.

\subsubsection{Discrete Heat Equation Invariant Under Dilations}
\label{subsubsec3.3.3}

Let us now consider a five-point lattice that is also uniform and orthogonal.
We put
\begin{eqnarray}
\frac{u_{m, n+1}-u_{m, n}}{t_{m, n+1}-t_{m, n}} = \frac{u_{m+1n}-2u_{m,
n}+u_{m-1n}}{(x_{m+1, n}-x_{m, n})^2}\label{3.32}\\
x_{m+1n}-2x_{mn}+x_{m-1n} = 0 \quad\quad x_{mn+1}-x_{mn}  = 0\label{3.33}\\
t_{m+1n}-t_{mn} = 0 \quad\quad t_{m, n+1}-2t_{m, n}+t_{m, n-1} = 0.\label{3.34}
\end{eqnarray}
The variables $v_a$ that we shall substitute from eq.~(\ref{3.32}),
(\ref{3.33}) and (\ref{3.34}) are $x_{m+1, n}$, $t_{m+1, n}$, $x_{m, n+1}$,
$t_{m, n+1}$ and $u_{m, n+1}$. Applying $\pr X$ to eq.~(\ref{3.33}) we obtain
\begin{eqnarray}
&&\xi(2x_{mn}-x_{m-1, n}, t_{mn}, u_{m+1, n}) - 2\xi(x_{mn}, t_{mn}, u_{mn})
\nonumber\\
&&\hspace{1.55in} + \xi(x_{m-1, n}, t_{mn}, u_{m-1,n}) = 0\label{3.35}\\
&&\xi(x_{mn}, 2t_{m, n}-t_{mn-1}, u_{m, n+1}) - \xi(x_{mn}, t_{mn}, u_{mn}) =
0.\label{3.36}
\end{eqnarray}
In eq.~(\ref{3.36}) $u_{mn}$ and $u_{m, n+1}$ are independent. Differentiating
with respect to $u_{mn}$ we find $\partial\xi_{mn}/\partial u_{mn} = 0$ and
hence $\xi$ does not depend on $u$. Differentiating (\ref{3.36}) with respect
to $t_{mn-1}$ we obtain $\partial\xi_{m, n+1}/\partial t_{m, n+1} = 0$. Thus,
$\xi$ depends on $x$ alone. Eq.~(\ref{3.35}) can then be solved and we find
that $\xi$ is linear in $x$. Applying $\pr X$ to eq.~(\ref{3.34}) we obtain
similar results for $\tau(x, t, u)$. Finally, invariance of the lattice
equations (\ref{3.33}) and (\ref{3.34}) implies:
\begin{equation}\label{3.37}
\xi = ax + b, \quad \tau = ct + d.
\end{equation}
Let us now apply $\pr X$ to eq.~(\ref{3.32}). We obtain, after using the
P$\Delta$S (\ref{3.32}) - (\ref{3.34})
\begin{eqnarray}
&&\lefteqn{\frac{\phi_{mn+1}-\phi_{mn}}{t_{m, n+1}-t_{m, n}} -
\frac{\phi_{m+1n}-2\phi_{mn}+\phi_{m-1n}}{(x_{m+1, n}-x_{m, n})^2}}\nonumber\\ 
&&\hspace{1.2in}+ (2a-c) \frac{u_{m+1, n}-2u_{mn}+u_{m-1,
n}}{(x_{m+1n}-x_{mn})^2} = 0.\label{3.38}
\end{eqnarray}
Notice that $u_{m, n+1}$ (and hence $\phi_{m, n+1}$) depends on $u_{m+1, n}$ and
$u_{m-1, n}$, whereas all terms in eq.~(\ref{3.38}) depend on at most one of
these quantities. Taking the second derivative $\partial u_{m+1, n}\partial
u_{m-1, n}$ of eq.~(\ref{3.38}), we find
\begin{equation}\label{3.39}
\frac{\partial^2\phi_{m, n+1}}{\partial u_{m, n+1}} = 0, \mbox{ i.e. } \phi =
A(x, t)u + B(x, t).
\end{equation}
We substitute this expression back into (\ref{3.38}) and find
\begin{equation}\label{3.40}
A(x, t) = A_0 = \const
\end{equation}
and see that $B(x, t)$ must satisfy the system (\ref{3.32})--(\ref{3.34}).
Moreover, we find $c = 2a$ in eq.~(\ref{3.37}). Finally, the symmetry algebra
has the basis
\begin{eqnarray}
P_0 &=& \partial_t, \quad P_1 = \partial_x, \quad W = u\partial_u, \quad
D = x\partial_x + 2t\partial_t\label{3.41}\\
S &=& S(x, t)\partial u.
\end{eqnarray}
Thus, dilational invariance is recovered, not however Galilei invariance.
Other symmetries can be recovered on other lattices.

\subsection{Lorentz Invariant Difference Schemes}\label{subsec3.4}

\subsubsection{The Continuous Case}\label{subsubsec3.4.1}

Let us consider the PDE
\begin{equation}\label{3.42}
u_{xx} - u_{tt} = 4f(u).
\end{equation}
Eq.~(\ref{3.42}) is invariant under the Poincar\'e group of $1+1$ dimensional
Minkowski space for any function $f(u)$. Its Lie algebra is represented by
\begin{equation}\label{3.43}
P_0 = \partial_t, \quad P_1 = \partial_x, \quad L =
t\partial_x + x\partial_t.
\end{equation}
For specific interactions $f(u)$ the symmetry algebra may be larger, in
particular for $f = e^u$, $f = u^N$, or $f = \alpha u + \beta$.

Before presenting a discrete version of eq.~(\ref{3.42}), we find it convenient
to pass over to light cone coordinates
\begin{equation}\label{3.44}
y = x + t, \quad z = x - t
\end{equation}
in which eq.~(\ref{3.42}) is rewritten as
\begin{equation}\label{3.45}
u_{yz} = f(u)
\end{equation}
and the Poincar\'e symmetry algebra (\ref{3.43}) is
\begin{equation}\label{3.46}
P_1 = \partial_y, \quad P_2 = \partial_z, \quad L = y\partial_y -
z\partial_z.
\end{equation}

\subsubsection{A Discrete Lorentz Invariant Scheme}\label{subsub3.4.2}

A particular P$\Delta$S that has eq.~(\ref{3.45}) as its continuous limit is
\begin{eqnarray}
\frac{u_{m+1n+1}-u_{mn+1}-u_{m+1n}+u_{mn}}{(y_{m+1n}-y_{mn})(z_{mn+1}-z_{mn})}
= f(u_{mn})\label{3.47}\\
y_{m+1n} - 2y_{mn} + y_{m-1n} = 0 \quad y_{mn+1} - y_{mn} = 0\label{3.48}\\
z_{m+1n} - z_{mn} = 0 \quad z_{mn+1} - 2z_{mn} + z_{mn-1} = 0.\label{3.49}
\end{eqnarray}
To find the Lie point symmetries of this difference scheme, we put
\begin{equation}\label{3.50}
X = \eta(y, z, u)\partial_y + \xi(y, z, u)\partial_z + \phi(y, z,
u)\partial_u.
\end{equation}
We apply the prolonged vector field $\pr\widehat x$ first to eq.~(\ref{3.48})
and (\ref{3.49}), eliminate $y_{m+1n}$, $y_{m, n+1}$, $z_{m+1n}$ and $z_{m,
n+1}$, using the system (\ref{3.48}), (\ref{3.49}) and notice that all $u_{ik}$
that figure in the obtained equations for $\eta_{ik}$ and $\xi_{ik}$ are
independent.

The result that we obtain is that $\eta$ and $\xi$ must be independent of $u$
and linear in $y$ and $z$, respectively. Finally we obtain
\begin{equation}\label{3.51}
\xi = \alpha y + \gamma, \quad \eta = \beta z + \delta
\end{equation}
($\alpha, \dots, \delta$ are constants).
Invariance of eq.~(\ref{3.47}) implies that the coefficient $\phi$ in the
vector field (\ref{3.50}) must be linear in $u$ and moreover have the form
\begin{equation}\label{3.52a}
\phi = Au + B(y, z)
\end{equation}
where $A$ is a constant. Taking (\ref{3.51}) and (\ref{3.52a}) into account and
applying $\pr X$ to eq.~(\ref{3.47}), we obtain
\begin{eqnarray}
(A-\alpha-\beta)f(u_{mn}) + \frac{B_{m+1n+1}-B_{mn+1}-B_{m+1n}+B_{mn}}
{(y_{m+1, n}-y_{mn})(z_{mn+1}-z_{mn})}\nonumber\\
= (Au_{mn}+B_{mn})f'(u_{mn}).\label{3.52b}
\end{eqnarray}
Differentiating eq.~(\ref{3.52b}) with respect to $u_{mn}$ we finally obtain the
following determining equation:
\begin{equation}\label{3.53}
(\alpha+\beta) \frac{df}{du_{mn}} + [Au_{mn}+B(y_{mn}, z_{mn})] \frac{d^2f}
{du_{mn}^2} = 0.
\end{equation}
For $f(u_{mn})$ arbitrary, we find $\beta = - \alpha$, $A = B = 0$. Thus for
arbitrary $f(u)$ the scheme (\ref{3.47})--(\ref{3.48}) has the same
symmetries as its continuous limit. The point symmetry algebra is given by eq.~
(\ref{3.46}), i.e.\ it generates, translations and Lorentz transformations.

Now let us find special cases of $f(u)$ when further symmetries exist. That
means that eq.~(\ref{3.53}) must be solved in a nontrivial manner. Let us
restrict ourselves to the case when the interaction is nonlinear, i.e.
\begin{equation}\label{3.54}
\frac{d^2}{du_{mn}^2}f(u_{m, n}) \neq 0.
\end{equation}
Then we must have
\begin{equation}\label{3.55}
B(y_{mn}, z_{mn}) = B = \const.
\end{equation}
The equation to be solved for $f(u)$ is actually eq.~(\ref{3.52b}) which
simplifies to
\begin{equation}\label{3.56}
(Au+B)f'(u) = (A-\alpha-\beta)f(u).
\end{equation}
For $A\neq 0$ the solution of eq.~(\ref{3.56}) is
\begin{equation}\label{3.57}
f = f_0u^p
\end{equation}
and the symmetry is
\begin{equation}\label{3.58}
D_1 = y\partial_y + z\partial_z - \frac{2}{p-1}u\partial u.
\end{equation}
For $A = 0$, $B \neq 0$ we obtain
\begin{equation}\label{3.59}
f = f_0e^{pu}
\end{equation}
and the additional symmetry is
\begin{equation}\label{3.60}
D_2 = y\partial_y + z\partial_z - 2\partial_u.
\end{equation}

Thus, for nonlinear interactions $f(u)$, $f'' \neq 0$, the P$\Delta$S
(\ref{3.47})--(\ref{3.49}) has exactly the same point symmetries as
its continuous limit (\ref{3.45}).

The linear case
\begin{equation}\label{3.61}
f(u) = Ru + T
\end{equation}
is different. The PDE (\ref{3.45}) in this case is conformally invariant. This
infinite dimensional symmetry algebra is not present for the discrete case
considered in this section.

\section{Symmetries of Discrete Dynamical Systems}\label{sec4}

\subsection{General Formalism}\label{subsec4.1}

In this chapter we shall discuss differential-difference equations on a fixed
one-dimensional lattice. Thus, time $t$ will be a continuous variable, $n \in
\mathbb Z$ a discrete one. We will be modeling discrete monoatomic or diatomic
molecular chains with equally spaced rest positions. The individual atoms will
be vibrating around their rest positions. For monoatomic chains the actual
position of the $n$-th atom is described by one continuous variable $u_n(t)$.
For diatomic atoms there will be two such functions, $u_n(t)$ and $v_n(t)$.

Only nearest neighbour interaction will be considered. The interaction are
described by a priori unspecified functions. Our aim is to classify these
functions according to their symmetries.

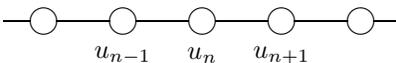
\begin{figure}
\begin{center}
\setlength{\unitlength}{2pt}
\begin{picture}(75,11)(0,-6)
\put(0,2.5){\line(1,0){5}}
\put(7.5,2.5){\circle{5}}
\put(10,2.5){\line(1,0){10}}
\put(22.5,2.5){\circle{5}}
\put(22.5,-4){\makebox(0,0){$u_{n-1}$}}
\put(25,2.5){\line(1,0){10}}
\put(37.5,2.5){\circle{5}}
\put(37.5,-4){\makebox(0,0){$u_{n}$}}
\put(40,2.5){\line(1,0){10}}
\put(52.5,2.5){\circle{5}}
\put(52.5,-4){\makebox(0,0){$u_{n+1}$}}
\put(55,2.5){\line(1,0){10}}
\put(67.5,2.5){\circle{5}}
\put(70,2.5){\line(1,0){5}}
\end{picture}
\end{center}
\caption{A monoatomic chain}
\label{fig4.1}
\end{figure}

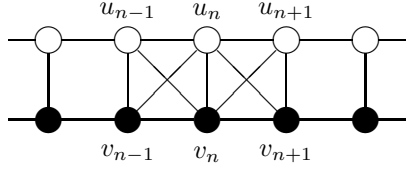
\begin{figure}
\begin{center}
\setlength{\unitlength}{2pt}
\begin{picture}(75,30)(0,-6)
\put(0,2.5){\line(1,0){5}}
\put(7.5,2.5){\circle*{5}}
\put(10,2.5){\line(1,0){10}}
\put(22.5,2.5){\circle*{5}}
\put(22.5,-4){\makebox(0,0){$v_{n-1}$}}
\put(25,2.5){\line(1,0){10}}
\put(37.5,2.5){\circle*{5}}
\put(37.5,-4){\makebox(0,0){$v_{n}$}}
\put(40,2.5){\line(1,0){10}}
\put(52.5,2.5){\circle*{5}}
\put(52.5,-4){\makebox(0,0){$v_{n+1}$}}
\put(55,2.5){\line(1,0){10}}
\put(67.5,2.5){\circle*{5}}
\put(70,2.5){\line(1,0){5}}
\put(0,17.5){\line(1,0){5}}
\put(7.5,17.5){\circle{5}}
\put(10,17.5){\line(1,0){10}}
\put(22.5,17.5){\circle{5}}
\put(22.5,23){\makebox(0,0){$u_{n-1}$}}
\put(25,17.5){\line(1,0){10}}
\put(37.5,17.5){\circle{5}}
\put(37.5,23){\makebox(0,0){$u_{n}$}}
\put(40,17.5){\line(1,0){10}}
\put(52.5,17.5){\circle{5}}
\put(52.5,23){\makebox(0,0){$u_{n+1}$}}
\put(55,17.5){\line(1,0){10}}
\put(67.5,17.5){\circle{5}}
\put(70,17.5){\line(1,0){5}}
\multiput(7.5,5)(15,0){5}{\line(0,1){10}}
\multiput(22.5,2.5)(15,0){2}{\line(1,1){13}}
\multiput(37.5,2.5)(15,0){2}{\line(-1,1){13}}
\end{picture}
\end{center}
\caption{A diatomic molecule with two types of atoms on parallel chains}
\label{fig4.2}
\end{figure}

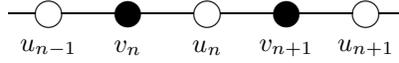
\begin{figure}
\begin{center}
\setlength{\unitlength}{2pt}
\begin{picture}(75,11)
\put(0,2.5){\line(1,0){5}}
\put(7.5,2.5){\circle{5}}
\put(7.5,-4){\makebox(0,0){$u_{n-1}$}}
\put(10,2.5){\line(1,0){10}}
\put(22.5,2.5){\circle*{5}}
\put(22.5,-4){\makebox(0,0){$v_{n}$}}
\put(25,2.5){\line(1,0){10}}
\put(37.5,2.5){\circle{5}}
\put(37.5,-4){\makebox(0,0){$u_{n}$}}
\put(40,2.5){\line(1,0){10}}
\put(52.5,2.5){\circle*{5}}
\put(52.5,-4){\makebox(0,0){$v_{n+1}$}}
\put(55,2.5){\line(1,0){10}}
\put(67.5,2.5){\circle{5}}
\put(67.5,-4){\makebox(0,0){$u_{n+1}$}}
\put(70,2.5){\line(1,0){5}}
\end{picture}
\end{center}
\caption{A diatomic molecule with two types of atoms alternating along one chain}
\label{fig4.3}
\end{figure}

Three different models have been studied \cite{ref21,ref22,ref23}. They
correspond to Fig.~\ref{fig4.1}, \ref{fig4.2} and \ref{fig4.3}, respectively.

The model illustrated on Fig.~\ref{fig4.1} corresponds to the equation
\cite{ref21}
\begin{equation}\label{4.1}
\ddot u_n(t) - F_n\big(t, u_{n-1}(t), u_n(t), u_{n+1}(t)\big) = 0.
\end{equation}
Fig.~\ref{fig4.2} could correspond to a very primitive model of the DNA
molecule. The equations are \cite{ref22}
\begin{eqnarray}
\ddot u_n & = & F_n\big(t, u_{n-1}(t), u_n(t), u_{n+1}(t), v_{n-1}(t), v_n(t),
v_{n+1}(t)\big) = 0\nonumber\\
\ddot v_n & = & G_n\big(t, u_{n-1}(t), u_n(t), u_{n+1}(t), v_{n-1}(t), v_n(t),
v_{n+1}(t)\big) = 0.\label{4.2}
\end{eqnarray}

The model corresponding to Fig.~\ref{fig4.3} already took translational and
Galilei invariance into account, so the equations were
\begin{eqnarray}
\ddot u_n & = & F_n(\xi_n, t) + G_n(\eta_{n-1}, t)\nonumber\\
\ddot v_n & = & K_n(\xi_n, t) + P_n(\eta_{n-1}, t)\label{4.3}\\
\xi_n & = & y_n - x_n, \quad \eta_n = x_{n+1} - y_n.\nonumber
\end{eqnarray}

Dissipation was ignored in all three cases, so no first derivatives are present.

In these lectures we shall only treat the case (\ref{4.1}). The lattice is
fixed, i.e.\ it is given by the relation
\begin{equation}\label{4.4}
x_n = hn + x_0
\end{equation}
with $h$ and $x_0$ given constants. With no loss of generality we can choose $h
= 1$, $x_0 = 0$, so that we have $x_n = n$.

Our aim is to find all functions $F_n$ for which eq.~(\ref{4.1}) allows a
nontrivial group of local Lie point transformations. We shall also assume that
the interaction is nonlinear and that it does indeed couple neighbouring states.

Let us sum up the conditions imposed on the model (\ref{4.1}) and on the
symmetry studies.
\begin{enumerate}
\item[1.] The lattice is fixed and regular $(x_n = n)$.

\item[2.] The interaction involves nearest neighbours only, is nonlinear and
coupled, i.e.
\begin{equation}\label{4.5}
\frac{\partial^2F_n}{\partial u_i\partial u_k} \neq 0, \quad \frac{\partial
F_n}{\partial u_{n-1}} \neq 0, \quad \frac{\partial F_n}{\partial u_{n+1}} \neq
0.
\end{equation}

\item[3.] We consider point symmetries only. Since the lattice is fixed, the
transformations are generated by vector fields of the form \cite{ref49}
\begin{equation}\label{4.6}
\widehat X = \tau(t)\partial_t + \phi_n(t, u_n)\partial u_n.
\end{equation}
We also assume that $\tau(t)$ is an analytical function of $t$ and $\phi_n(t,
u_n)$ is also analytic as a function of $t$ and $u_n$.
\end{enumerate}

The symmetry algorithm is the usual one, namely

\begin{equation}\label{4.7}
\pr \widehat XE_n\mid_{E_n = 0} = 0.
\end{equation}

The prolongation in eq.~(\ref{4.7}) involves a prolongation to $t$-derivatives
$\dot u_n$ and $\ddot u_n$, and to all values of $n$ figuring in
eq.~(\ref{4.1}), i.e. $n \pm 1$.

The terms in the prolongation that we actually need are
\begin{equation}\label{4.8}
\pr^{(2)}X = \tau\partial_t + \sum^{n+1}_{k=n-1} \phi_k(t, u_k)\partial_{u_k} +
\phi^{tt}_n \partial_{\ddot u_n}.
\end{equation}
The coefficient $\phi^{tt}_n$ is calculated using the formulas of
Chapter~\ref{sec1} (or e.g.\ Ref.~\cite{ref1}). We have
\begin{equation}\label{4.9}
\phi^{tt}_n = D^2_t\phi_n - (D^2_t\tau)u_n -2(D_t\tau)\ddot u_n.
\end{equation}
Applying $\pr^{(2)}X$ to eq.~(\ref{4.1}) and replacing
$\ddot u$ from that equation, we get an expression involving $(\dot u_n)^3$,
$(\dot u_n)^2$, $(\dot u_n)^1$ and $(\dot u_n)^0$. The coefficients of all of
these terms must vanish separately. The first three of these equations do not
depend on $F_n$ and can be solved easily. They imply
\begin{eqnarray}\label{4.10}
\phi_n(t, u_n) = \left(\frac 12\dot{\tau}(t)+a_n\right)u_n + \beta_n(t), \quad
\tau = \tau(t), \quad \dot a_n = 0.
\end{eqnarray}
The remaining determining equation is
\begin{eqnarray}
\lefteqn{\frac 12 \tau_{ttt} u_n + \ddot{\beta}_n + \left(a_n+\frac
32\dot{\tau}\right) F_n}\nonumber\\
&&\hspace{0.50in} - \tau F_{n, t} - \sum_{\alpha}\left[\left(\frac
12\dot{\tau}+a_{\alpha}\right) u_{\alpha}+\beta_{\alpha}\right] F_{n,
u_{\alpha}}  = 0,\label{4.11}
\end{eqnarray}
and the vector fields realizing the symmetry algebra are
\begin{equation}\label{4.12}
\widehat X = \tau(t)\partial_t + \left[\left(\frac 12\dot{\tau}(t)+a_n\right)
u_n+\beta_n(t)\right]\partial u_n.
\end{equation}

Since we are classifying the interactions $F_n$, we must decide which functions
$F_n$ will be considered to be equivalent. To do this we introduce a group of
``allowed transformations'', or a ``classifying group''. We define this to be a
group of fiber preserving point transformations
\begin{equation}\label{4.13}
u_n(t) = \Omega_n(\tilde u_n(\tilde t), \tilde t, g), \quad \tilde t = \tilde
t(t, g), \quad \tilde n = n,
\end{equation}
taking eq.~(\ref{4.1}) into an equation of the same form
\begin{equation}\label{4.14}
\ddot{\tilde u}_n(\tilde t) = \tilde F_n\big(\tilde t, \tilde
u_{n-1}(\tilde t), \tilde u_n(\tilde t), \tilde u_{n+1}(\tilde t)\big) = 0.
\end{equation}
That is, the allowed transformations can change the function $F_n$ (as opposed
to symmetry transformations), but cannot introduce first derivatives, nor other
than nearest neighbour terms. These conditions narrow down the transformations
(\ref{4.13}) to linear ones of the form
\begin{eqnarray}
u_n(t) & = & \frac{A_n}{\sqrt{\tilde t_t}}\tilde u_n(\tilde t) + B_n(t), \quad
\tilde t = \tilde t(t)\nonumber\\
A_{n, t} & = & 0, \quad \tilde t_t \neq 0, \quad A_n \neq 0, \quad \tilde n =
n\label{4.15}.
\end{eqnarray}
Eq.~(\ref{4.1}) is transformed into
\begin{eqnarray}
\lefteqn{\tilde u_{n, \tilde t\tilde t} = A^{-1}_n(\tilde t_t)^{-3/2}
        \biggl\{
F_n(t, u_{n-1}, u_n, u_{n+1})}\nonumber\\
&\hspace{1in}& \quad\quad+\biggl[-\frac 34A_n(\tilde t_t)^{-5/2}(\tilde
t_{tt})^2 \nonumber\\ 
&\hspace{1in}& \qquad\qquad\qquad +\frac{A_n}{2}(\tilde t_t)^{-3/2}\tilde t_{ttt}\biggr]\tilde
u_n(\tilde t)- B_{n,
tt}\biggr\}.\label{4.16}
\end{eqnarray}
The transformed vector field (\ref{4.12}) is
\begin{eqnarray}
\lefteqn{\widehat X = \tau(t) \tilde t_t(t)\partial_{\tilde t} + \left\{
\left[\frac 12\big(\tau(t)\tilde t_t\big)_{\tilde t}+a_n\right]\tilde
u_n\right\}}\nonumber\\
&\hspace{0.5in}& + \left.(\tilde t_t)^{1/2}A^{-1}_n\left[\left(\frac
12\tau_t+a_n\right)B_n + \beta_n - \tau B_{n, t}\right]\right\}\partial\tilde
u_n.\label{4.17} 
\end{eqnarray}
In eq.~(\ref{4.16}) and (\ref{4.17}) $\tau(t)$, $a_n$, $\beta_n$ and $F_n$
are given, whereas $\tilde t(t)$, $A_n$ and $B_n(t)$ are ours to choose. We use
these quantities to simplify the vector field $\widehat X$.

Our classification strategy is the following one. We first classify
one-\hspace*{0pt}dimensional subalgebras. Thus, we have one vector field of the form
(\ref{4.12}). If $\tau(t)$ satisfies $\tau(t) \neq 0$ in some open interval, we
use $\tilde t(t)$ to normalize $\tau(t) = 1$ and $B_n(t)$ to transform
$\beta_n(t)$ into $\beta_n(t) = 0$. If we have $\tau(t) = 0$, $a_n \neq 0$,
we use $B_n(t)$ to annul $\beta_n(t)$. The last possibility is $\tau(t)
= 0$, $a_n = 0$, $\beta_n(t) \neq 0$. Then we cannot simplify further. The same
transformations will also simplify the determining equation~(\ref{4.11}) and we
can, in each case, solve it for the interaction $F_n(t, u_{n-1}, u_n, u_{n+1})$.

Once all interactions allowing one dimensional symmetry algebras are
determined, we proceed further structurally. We first find all Abelian symmetry
groups and the corresponding interactions allowing them. We run through our
list of one-dimensional algebras and take them in an already establlished
``canonical'' form. Let us call this element $X_1$ (in each case). We then find
all elements $X$ of the form (\ref{4.12}) that satisfy $[X_1, X] = 0$. We
classify the obtained operators $X$  under the action of a subgroup of the
group of allowed transformations, namely the isotropy group of $X_1$ (the group
that leaves the subalgebra $X_1$ invariant). For each Abelian group we find the
invariant interaction.

>From Abelian symmetry algebras we proceed to nilpotent ones, then to solvable
ones and finally to nonsolvable ones. These can be semisimple, or they may have
a nontrivial Levi decomposition.

All details can be found in the original article \cite{ref21}, here we shall
present the main results.

\subsection{One-Dimensional Symmetry Algebras}\label{subsec4.2}

Three classes of one-dimensional symmetry algebras exist. Together with their
invariant interactions, they can be represented by
\begin{eqnarray}
A_{1, 1} \qquad && X = \partial_t + a_nu_n\partial
u_n\nonumber\\ 
&& F_n(t, u_k) = f_n(\xi_{n-1}, \xi_n,
\xi_{n+1})e^{a_nt}\label{4.18}\\ && \xi_k = u_ke^{-a_kt}, \quad k = n-1, n,
n+1.\nonumber\\
&& \nonumber\\
A_{1, 2} \qquad &&
X = a_nu_n\partial u_n\nonumber\\
&& F_n(t, u_k) = u_nf_n(t, \xi_{n-1}, \xi_{n+1})\label{4.19}\\
&& \xi_{n\pm 1} = u^{a_n}_{n\pm 1} u^{a_{n\pm 1}}_n.\nonumber\\
&& \nonumber\\
A_{1, 3} \qquad && X = \beta_n(t)\partial u_n\nonumber\\
&& F_n(t, u_k) = \frac{\ddot{\beta}_n}{\beta_n}u_n + f_n(t, \xi_{n-1},
\xi_{n+1})\nonumber\\
&& \xi_{n\pm 1} = \beta_n(t)u_{n\pm 1} - \beta_{n\pm 1}(t) u_n.\label{4.20}
\end{eqnarray}
We see that the existence of a one-dimensional Lie algebra implies that the
interaction $F$ is an arbitrary function of three variables, rather than the
original four. The actual form of the interaction in eq.~(\ref{4.18}),
(\ref{4.19}) and (\ref{4.20}) was obtained by solving eq.~(\ref{4.11}), once
the canonical form of vector field $X$ in eq.~(\ref{4.18}),
(\ref{4.19}), or (\ref{4.20}) was taken into account.

\subsection{Abelian  Lie Algebras of Dimension $N \geq 2$}\label{subsec4.3}

Without proof we state several theorems.

\begin{theorem}\label{thm4.1}
An Abelian symmetry algebra of eq.~\textup{(\ref{4.1})} can have dimension
$N$ satisfying $1 \leq N \leq 4$.
\end{theorem}

\noindent\emph{Comment}: For $N = 1$ these are the algebras $A_{1, 1}$, $A_{1,
2}$ and $A_{1, 3}$ of eq.~(\ref{4.18}), (\ref{4.19}) and (\ref{4.20}).

\begin{theorem}\label{thm4.2}
Five distinct classes of interactions $F_n$ exist having symmetry algebras of
dimension $N = 2$. For four of them the interaction will involve an arbitrary
function of two variables, for the fifth $a$ function of three variables.
\end{theorem}

The five classes can be represented by the following algebras and interactions.
\begin{eqnarray}
A_{2, 1}: \quad && X_1 = \partial_t + a_{1n}u_n\partial u_n, \quad X_2 = a_
{2n}u_n\partial u_n\nonumber\\
&& F_n = u_nf_n(\xi_{n-1}, \xi_{n+1}), \quad a_{2n} \neq 0\label{4.21}\\
&&\xi_k = u^{a_{2n}}_k u^{-a_{2k}}_ne^{(a_{1n}a_{2k}-a_{1k}a_{2n})t}, \qquad k =
n \pm 1\nonumber\\
&&\nonumber\\
A_{2, 2}: \quad && X_1 = \partial_t + a_nu_n\partial u_n \quad X_2 = e^{a_nt}
\partial u_n\nonumber\\
&& F_n = a^2_nu_n + e^{a_nt}f_n(\xi_{n-1}, \xi_{n+1})\label{4.22}\\
&& \xi_k = u_ke^{-a_kt} - u_n e^{-a_nt}, \quad k = n \pm 1\nonumber\\
&& \nonumber\\
A_{2, 3}: \quad && X_1 = a_{1n}u_n\partial u_n \quad X_2 = a_{2n}u_n\partial 
u_n\nonumber\\
&& F_n = u_nf_n(t, \xi)\label{4.23}\\
&& \xi = u^{\alpha_{n+1, n}}_{n-1}u^{\alpha_{n-1, n+1}}_n u^{\alpha_{n, n-1}}_
{n+1}\nonumber\\
&& \alpha_{kl} = a_{1k}a_{2l}-a_{1l}a_{2k} \neq 0\nonumber\\
&&\nonumber\\
A_{2, 4}: \quad && X_1 = \beta_{1, n}(t)\partial u_n, \quad X_2 = \beta_{2n}(t)
\partial u_n\nonumber\\
&& \quad\quad\quad \beta_{1n}\beta_{2n+1} - \beta_{1n+1}\beta_{2n} \neq
0\label{4.24}\\
&& F_n = \frac{(\beta_{1n}\ddot{\beta}_{2n}-\ddot{\beta}_{1n}\beta_{2n}) u_{n+1}
- (\beta_{1n+1}\ddot{\beta}_{2n} - \ddot{\beta}_{1n}\beta_{2n+1})}{\beta_{1n}
\beta_{2n+1} - \beta_{1n+1}\beta_{2n}}\nonumber\\
&& \quad\quad\quad + f_n(t, \xi)\nonumber\\
&& \xi = (\beta_{1n}\beta_{2n+1} - \beta_{1n+1}\beta_{2n}) u_{n-1} +
(\beta_{1n+1}\beta_{2n-1} - \beta_{1n-1}\beta_{2n+1})u_n\nonumber\\
&&\quad\quad\quad + (\beta_{1n-1}\beta_{2n} -
\beta_{1n}\beta_{2n-1})u_{n+1}\nonumber\\
&&\nonumber\\
A_{2, 5}: \quad && X_1 = \partial u_n, \quad X_2 = t\partial u_n\nonumber\\
&& F_n = f_n(t, \xi_{n-1}, \xi_{n+1}), \quad \xi_k = u_k - u_n, \quad k = n \pm
1Ë\label{4.25}
\end{eqnarray}

The algebra $A_{2, 5}$ is of particular physical significance since $X_1$ and
$X_2$ in eq.~(\ref{4.25}) correspond to translation and Galilei invariance
for the considered chain. Unless we are considering a molecular chain in some
external field, or unless some external geometry is imposed, the symmetry
algebra $A_{2, 5}$ should always be present, possibly as a subalgebra of a
larger symmetry algebra.

\begin{theorem}\label{thm4.3}
Precisely four classes of three-dimensional symmetry algebras exist. Only one
of them contains the $A_{2, 5}$ subalgebra and can be presented as
\begin{eqnarray}
A_{3, 4} \quad && X_1 = \partial u_n, \quad X_2 = t\partial u_n,\nonumber\\ 
&& X_3 = \beta_n(t)\partial u_n,\quad \beta_{n+1} \neq \beta_n, \,
\ddot{\beta}_n \neq 0.\label{4.26}
\end{eqnarray}
The invariant interaction is
\begin{eqnarray}
F_n & = & \frac{\ddot{\beta}_n}{\beta_{n+1}-\beta_n} (u_{n+1}-u_n) + f_n(t,
\xi),\label{4.27}\\
\xi & = & (\beta_n-\beta_{n+1})u_{n-1} + (\beta_{n+1}-\beta_{n-1})u_n +
(\beta_{n-1}-\beta_n) u_{n+1}.\label{4.28}
\end{eqnarray}
\end{theorem}
For $A_{3, 1}$, $A_{3, 2}$ and $A_{3, 4}$ see the original article \cite{ref21}.

\begin{theorem}\label{thm4.4}
There exist precisely two classes of interactions $F_n$ in
eq.~\textup{(\ref{4.1})} satisfying conditions \textup{(\ref{4.5})},
allowing four-dimensional symmetry algebras. Only one of them contains the
subalgebra $A_{2, 5}$ and is represented by the following.
\begin{eqnarray}
A_{4, 1} \quad && F_n = \frac{B_n(t)\gamma_n}{\gamma_n-\gamma_{n+1}}
(u_n-u_{n+1}) + f_n(t, \xi), f_{n, \xi\xi} \neq 0\nonumber\\
&& X_1 = \partial u_n, \quad X_2 = t\partial u_n, \quad X_3 = \psi_1(t) \gamma_n
\partial u_n,\label{4.29}\\
&& \quad\quad\quad X_4 = \psi_2(t)\gamma_n\partial u_n\nonumber\\
&& \gamma_{n+1} \neq \gamma_n, \quad \dot{\gamma}_n =  0, \quad \psi_1\dot{\psi}
_2 - \dot{\psi}_1\psi_2 = \const \neq 0\nonumber
\end{eqnarray}
with $\xi$ as in eq.~(\ref{4.28}) and $\psi_1$, $\psi_2$ satisfying
\[
\ddot{\psi}_i - B(t)\psi_i = 0, \quad i = 1, 2
\]
\end{theorem}

\subsection{Some Results on the Structure of Lie Algebras}\label{subsec4.4}

Let us recall some basic properties of finite-dimensional Lie algebras.
Consider a Lie algebra $L \sim \{X_1, X_2, \dots, X_n\}$, where the elements
$X_i$ form a basis. To each algebra $L$ one associates two series of
subalgebras.

\emph{The derived series} consist of the algebras
\begin{eqnarray}
L^0 \equiv L, \quad L^1 \equiv DL = [L, L], \quad L^2 \equiv D^2L =
[DL, DL], \dots\nonumber\\ 
L^N \equiv D^NL = [D^{N-1}L, D^{N-1}L].\label{4.30}
\end{eqnarray}
The algebra of commutators $DL$ is called the \emph{derived algebra}. If we
have $DL = L$, the algebra $L$ is called \emph{perfect}. If an integer $N$
exists for which we have $D^NL = \{0\}$, the algebra $L$ is called
\emph{solvable}.

The \emph{central series} consist of the algebras
\begin{equation}\label{4.31}
L_0 \equiv L, \quad L_1 = L^1  = [L, L], \quad L_2 = [L, L_1], \dots L_N = [L,
L_{N-1}], \dots
\end{equation}

If there exists an integer $N$ for which we have $L_N = \{0\}$, the algebra $L$
is called \emph{nilpotent}. Clearly, every nilpotent algebra is solvable, but
the converse is not true.

Let us consider two examples
\begin{enumerate}
\item[1.] The Lie algebra of the Euclidean group of a plane: $e(2) \sim
\{L_3, P_1, P_2\}$. The commmutation relations are
\begin{equation}\label{4.32}
[L_3, P_1] = P_2, \quad [L_3, P_2] = - P_1, \quad [P_1, P_2] = 0.
\end{equation}
The derived series is
\[
L = \{L_3, P_1, P_2\} \supset DL = \{P_1, P_2\}, \quad D^2L = \{0\}
\]
and the central series is
\[
L \supset L_1 = \{P_1, P_2\} = L_2 = L_3 = \dots
\]
Hence $e(2)$ is solvable but not nilpotent.

\item[2.] The Heisenberg algebra $H_1 \sim \{X_1, X_2, X_3\}$ where the basis
can be reallized e.g.\ by the derivative operator, the coordinate $x$ and the
identity 1:
\[
X_1 = \partial_x, X_2 = x, X_3 = 1.
\]
We have
\begin{equation}\label{4.33}
[X_1, X_2] = X_3, \quad [X_1, X_3] = [X_2, X_3] = 0
\end{equation}
and hence
\begin{eqnarray}
DL & = & \{X_3\}, \quad D^2L = 0.\nonumber\\
L_1 & = & \{X_3\}, \quad L_2 = 0.\nonumber
\end{eqnarray}
We see that the Heisenberg algebra is nilpotent (and solvable).
\end{enumerate}

An Abelian Lie algebra is of course also nilpotent.

We shall need some results concerning nilpotent Lie algebras (by nilpotent we
mean nilpotent non-Abelian).
\begin{enumerate}
\item[1.] Nilpotent Lie algebras always contain Abelian ideals.

\item[2.] All nilpotent Lie algebras contain the three-dimensional Heisenberg
algebra as a subalgebra.
\end{enumerate}

We shall also use some basic properties of solvable Lie algebras, where by
solvable we mean solvable, non-nilpotent.
\begin{enumerate}
\item[1.] Every solvable Lie algebra $L$ contains a unique maximal nilpotent
ideal called the nilradical $NR(L)$. The dimension of the nilradical satisfies
\begin{equation}\label{4.34}
\frac 12 \dim(L) \leq \dim NR(L) \leq \dim(L) - 1.
\end{equation}

\item[2.] If the nilradical $NR(L)$ is Abelian, then we can choose a basis for
$L$ in the form $\{X_1, \dots, X_n, Y_1, \dots, Y_m\}, \quad m \leq n$, with
commutation relations
\begin{equation}
[X_i, X_k] = 0, \quad [X_i, Y_k] = (A_k)_{ij}X_j, \quad 
[Y_i, Y_k] = C^l_{ik}X_l.\label{4.35}
\end{equation}
The matrices $A_k$ commute and are linearly nilindependent (i.e.\ no linear
combination of them is a nilpotent matrix).
\end{enumerate}

If a Lie algebra $L$ is not solvable, it can be simple, semisimple, or it may
have a nontrivial Levi decomposition \cite{ref15}. A \emph{simple} Lie algebra
$L$ has no nontrivial ideals, i.e.
\begin{equation}\label{4.36}
I\subseteq L, \quad [I, I] \subseteq I, \quad [L, I] \subseteq I
\end{equation}
implies $I \sim \{0\}$, or $I = L$.

\emph{A semisimple} Lie algebra $L$ is a direct sum of simple Lie algebras
$L_i$
\begin{equation}\label{4.37}
L \sim L_1 \oplus L_2 \oplus \dots \oplus L_p, \quad [L_i, L_k] = 0.
\end{equation}
If $L$ is not simple, semisimple, or solvable, then it allows a unique
\emph{Levi decomposition} into a semidirect sum
\begin{equation}\label{4.38}
L \sim S \semisum R, \quad [S, S] = S, \quad [R, R] \subset R, \quad [S, R]
\subseteq R
\end{equation}
where $S$ is semisimple and $R$ is solvable; $R$ is called the radical of $L$,
i.e.\ the maximal solvable ideal.

Let us now return to the symmetry classification of discrete dynamical systems.

\subsection{Nilpotent Non-Abelian Symmetry Algebras}\label{subsec4.5}

Since every nilpotent Lie algebra contains the three-dimensional Heisenberg
algebra, we start by constructing this algebra, $H_1 \sim \{X_1, X_2, X_3\}$.
The central element $X_3$ of eq.~(\ref{4.33}) is uniquely defined. We start
from this element, take it in one of the standard forms (\ref{4.18}),
(\ref{4.19}), or (\ref{4.20}), then construct the two complementary elements
$X_1$ and $X_2$. The result is that two inequivalent realizations of $H_1$,
exist namely:
\begin{eqnarray}
N_{3, 1}: \quad && X_1 = \partial_{u_n}, \quad X_2 = \partial_t, \quad X_3 =
t\partial_{u_n}\nonumber\\
&& F_n = f_n(\xi_{n+1}, \xi_{n-1}), \quad \xi_k = u_k - u_n, \quad k = n \pm
1\label{4.39}\\
&&\nonumber\\
N_{3, 2}: \quad && X_1 = e^{a_nt}\partial u_n, \quad X_2 = \partial_t +
a_nu_n\partial u_n\nonumber\\
&& X_3 = (t+\gamma_n)e^{a_nt}\partial u_n, \quad \dot a_n = 0, \quad
\dot{\gamma}_n = 0, \gamma_{n+1} \neq \gamma_n\nonumber\\
&& F_n = \frac{a^2_n(\gamma_{n+1}-\gamma_n)-2a_n}{\gamma_{n+1}-\gamma_n} u_n
\label{4.40}\\
&& \quad\quad\quad + \frac{2a_n}{\gamma_{n+1}-\gamma_n}
u_{n+1}e^{(a_n-a_{n+t})t}+ e^{a_nt}f_n(\xi)\nonumber\\
&& \xi = (\gamma_n-\gamma_{n+1})u_{n-1}e^{-a_{n-1}t} +
(\gamma_{n+1}-\gamma_{n-1}), u_ne^{-a_nt}\nonumber\\
&&\quad\quad\quad +(\gamma_{n-1}-\gamma_n)u_{n+1}e^{-a_{n+1}t}.\nonumber
\end{eqnarray}
Notice that $N_{3, 1}$ contains the physically important subalgebra $A_{2, 5}$.
whereas $N_{3, 2}$ does not.

Extending the algebras $N_{3, 1}$ and $N_{3, 2}$ by further elements, we find
that $N_{3, 1}$ gives rise to two five-dimensional nilpotent symmetry algebras
$N_{5, k}$  and $N_{3, 2}$ to a four-dimensional one $N_{4, 1}$.

Here we shall only give $N_{5, 1}$ and $N_{5, 2}$ which contain $N_{3, 1}$ and
hence
$A_{2, 5}$:
\begin{eqnarray}
N_{5, k}: \quad && X_1 = \partial u_n, \quad X_2 = t\partial u_n, \quad X_3 =
\biggl(\frac{(k-1)t^2}{2}+\gamma_n\biggr)\partial u_n,\nonumber\\
&& X_4 = \biggl(\frac{(k-1)t^3}{6} + \gamma_nt\biggr)\partial u_n, \quad X_5 =
\partial_t, \quad k = 1, 2\label{4.41}\\
&& F_n = \frac{2(k-1)}{\gamma_{n+1}-\gamma_n} (u_{n+1}-u_n) + f_n(\xi)\nonumber
\end{eqnarray}
with $\xi$ as in eq.~(\ref{4.40}).

\subsection{Solvable Symmetry Algebras with Non-Abelian Nilradicals}\label
{subsec4.6}

We already know all nilpotent symmetry algebras, so we can start from the
nilradical and extend it by further non-nilpotent elements. The result can be
stated as a Theeorem.

\begin{theorem}\label{thm4.5}
Seven classes of solvable symmetry algebras with non-Abelian nilradicals exist
for eq.~\textup{(\ref{4.1})}. Four of them have $N_{3, 1}$ as nilradical,
three have $N_{5, 1}$.
\end{theorem}

For $N_{3, 1}$ we can add just one further element $Y$, namely one of the
following
\begin{eqnarray}
SN_{4, 1}: \quad && Y = t\partial_t + \biggl(\frac 12+a\biggr)u_n\partial u_n,
\quad a \neq - \frac 12\nonumber\\
&& F_n = (u_{n+1}-u_n) e^{(a-3/2)/(a+1/2)}f_n(\xi)\label{4.42}\\
&&\nonumber\\
SN_{4, 2}: \quad && Y = t\partial_t + (2u_n+t^2)\partial u_n\nonumber\\
&& F_n = \ln(u_{n+1}-u_n) + f_n(\xi)\label{4.43}\\
&&\nonumber\\
SN_{4, 3}: \quad && Y = u_n\partial u_n\nonumber\\
&& F_n = (u_{n+1}-u_n)f_n(\xi).\label{4.44}
\end{eqnarray}
In all above cases we have
\begin{equation}\label{4.45}
\xi = \frac{u_{n-1}-u_n}{u_{n+1}-u_n}.
\end{equation}
\begin{eqnarray}
SN_{4, 4}: \quad && Y = t\partial_t + \gamma_n\partial u_n, \quad \gamma_{n+1}
\neq \gamma_n, \quad \dot{\gamma}_n = 0\nonumber\\
&& F_n = \exp\biggl(-2\frac{u_{n+1}-u_n}{\gamma_{n+1}-\gamma_n}\biggr)
f_n(\xi)\label{4.46}\\
&& \xi = (\gamma_n-\gamma_{n+1}) u_{n-1} + (\gamma_{n+1}-\gamma_{n-1}) u_n
\nonumber\\
&& \quad \quad \quad  +(\gamma_{n-1}-\gamma_n) u_{n+1}.\label{4.47}
\end{eqnarray}
For $N_{5, 1}$ we can also add at most one non-nilpotent element and we obtain
\begin{eqnarray}
SN_{6, 1}: \quad && Y = t\partial_t + \biggl(\frac 12+a\biggr) u_n\partial u_n
\nonumber\\
&& F_n = c_n\xi^{(a-3/2)/(a+1/2)}, \quad a \neq - \frac 12, \quad a \neq \frac
32
\label{4.48}\\
&&\nonumber\\
SN_{6, 2}: \quad && Y = t\partial_t +  [2u_n+(a+b\gamma_n)t^2] \partial u_n,
\quad a^2 + b^2 \neq 0\nonumber\\
&& F_n = c_n + (a+b\gamma_n)\ln\xi\label{4.49}\\
&&\nonumber\\
SN_ {6, 3}: \quad && Y = t\partial_t + \rho_n\partial u_n, \quad \rho_n \neq A
+ B\gamma_n , \quad \dot{\rho}_n \neq 0\nonumber\\
&& F_n = c_ne^{\zeta}\label{4.50}\\ 
&& \xi = \frac{-2\zeta}{(\gamma_n-\gamma_{n+1})
\rho_{n-1}+(\delta_{n+1}-\gamma_{n-1})\rho_n + (\gamma_{n-1}-\gamma_n)
\rho_{n+1}}.\nonumber
\end{eqnarray}
In all cases $\xi$ is as in eq.~(\ref{4.47}).

\subsection{Solvable Symmetry Algebras with Abelian Nilradicals}
\label{subsec4.7}

The results in this case are very rich. There exist 31 such symmetry algebras
and their dimensions satisfy $2 \leq d \leq 5$.

For all details and a full complete list of results we refer to the original
article. Here we give just one example of a five-dimensional Lie algebra with
$NR(L) = A_{4, 1}$.
\begin{eqnarray}
SA_{5, 1}: \quad && X_1 = \partial u_n, \quad X_2 = t\partial u_n, \quad X_3 =
e^t\gamma_n\partial u_n,\nonumber\\
&& \quad\quad\quad  X_4 = e^{-t}\gamma_n\partial u_n\nonumber\\
&& Y = \partial_t + au_n\partial u_n \quad a \neq 0, \quad \gamma_n \neq
\gamma_{n+1}, \quad \dot{\gamma}_n = 0\nonumber\\
&& F_n = \frac{\gamma_n(u_{n+1}-u_n)}{\gamma_{n+1}-\gamma_n} + e^{at}f_n(\xi)
\label{4.51}\\
&& \xi = [(\gamma_n-\gamma_{n+1}) u_{n-1} + (\gamma_{n+1}-\gamma_{n-1})u_n
\nonumber\\
&& \quad\quad\quad +(\gamma_{n-1}-\gamma_n) u_{n+1}]e^{-at}.\nonumber
\end{eqnarray}

\subsection{Nonsolvable Symmetry Algebras}\label{subsec4.8}

A Lie algebra that is not solvable must have simple subalgebra. The only simple
algebra that can be constructed out of vector fields of the form (\ref{4.12})
is $\Sl(2, \mathbb R)$. The algebra and the corresponding invariant interaction
can be represented as:
\begin{eqnarray}
NS_{3, 1}: \quad && X_1 = \partial_t, \quad X_2 = t\partial_t + \frac 12
u_n\partial u_n\nonumber\\
&& X_3 = t^2\partial_t + tu_n\partial u_n\label{4.52}\\
&& F_n = \frac{1}{u^3_n} f_n(\xi_{n-1}, \xi_{n+1}), \quad \xi_k =
\frac{u_k}{u_n}.\nonumber
\end{eqnarray}
This algebra can be further extended to a four, five or seven-dimensional
symmetry algebra. In two cases the algebra will have an $A_{2, 5}$ subalgebra,
namely

\noindent $NS_{5, 1}$: In addition to $X_1$, $X_2$, $X_3$ of (\ref{4.51}) we
have
\begin{eqnarray}
X_4 &=& \partial u_n, \quad X_5 = t\partial u_n\nonumber\\
F_n &=& (u_{n+1}-u_n)^{-3}f_n(\xi), \quad \xi = \frac{u_{n+1}-u_n}{u_{n-1}-u_n}.
\label{4.53}
\end{eqnarray}

\noindent $NS_{7, 1}$: The additional elements are
\begin{equation}\label{4.54}
\begin{array}{c}
X_n = \partial u_n, \quad X_5 = t\partial u_n, \quad X_6 = \gamma_n\partial
u_n, \quad X_7 = t\gamma_n\partial u_n\\
\gamma_{n+1} \neq \gamma_n, \quad \dot{\gamma}_n = 0.
\end{array}
\end{equation}

The invariant interaction is
\begin{eqnarray}
\lefteqn{F_n = s_n[(\gamma_n-\gamma_{n+1})u_{n-1} +
(\gamma_{n+1}-\gamma_{n-1})u_n}\nonumber\\
&\hspace{1in} & +
(\gamma_{n-1}-\gamma_n)u_{n+1}]^{-3},\quad \dot s_n = 0, \quad s_n \neq 0.
\end{eqnarray}

\subsection{Final Comments on the Classification}\label{subsec4.9}

Let us first of all sum up the discrete dynamical systems of the type
(\ref{4.1}) with the largest symmetry algebras

We  put
\begin{equation}
\xi = (\gamma_n-\gamma_{n+1})u_{n-1} + (\gamma_{n+1}-\gamma_{n-1})u_n +
(\gamma_{n-1}-\gamma_n)u_{n+1}
\end{equation}
and find this variable is involved in all cases with 7, or 6-dimensional
symmetry algebras.

The algebras and interactions are given in eq.~(\ref{4.54}), (\ref{4.48}),
(\ref{4.49}) and (\ref{4.50}), respectively.

A natural question to ask is: Where is the Toda lattice in this classification?
The Toda lattice is described by the equation
\begin{equation}\label{4.55}
u_{n, tt} = e^{u_{n-1}-u_n} - e^{u_n-u_{n+1}}.
\end{equation}
This equation is of the form (\ref{4.1}). It is integrable \cite{ref51} and has
many interesting properties. In our classification it appears as a special case
of the algebra $SN_{4, 4}$, i.e.
\begin{equation}\label{4.56}
\ddot u_n = \exp\biggl(-2\frac{u_{n+1}-u_n}{\gamma_{n+1}-\gamma_n}\biggr)
f_n(\xi),
\end{equation}
with
\begin{equation}\label{4.57}
f_n(\xi) = - 1 + e^{\xi/2}, \quad \gamma_n = 2n.
\end{equation}
Thus, its symmetry group is four-dimensional. We see that the Toda lattice is
not particularly distinguished by its point symmetries: other interactions have
larger symmetry groups. Even in the $SN_{4, 4}$ class two functions have to be
specialized (see eq.~(\ref{4.57}) to reduce (\ref{4.56}) to (\ref{4.55}).

\section{Generalized Point Symmetries of Linear and Linearizable Systems}
\label{sec5}

\subsection{Umbral Calculus}\label{subsec5.1}

In this chapter we take a different point of view than in the previous ones.
Instead of purely point symmetries, we shall consider a specific class of
generalized symmetries of difference equations that we shall call ``generalised
point symmetries''. They act simultaneously at several, or even infinitely many
points of a lattice, but they reduce to point symmetries of a differential
equation in the continuous limit.

The approach that we shall discuss here is at this stage applicable either to
linear difference equations, or to nonlinear equations that can be linearized
by a transformation of variables (not necessarily only point transformations).

The mathematical basis for this type of study is the so called ``umbral
calculus'' reviewed in recent books and articles by G.G. Rota and his
collaborators \cite{ref52,ref53,ref54}. Umbral calculus provides a unified basis
for studying symmetries of linear differential and difference equations.

First of all, let us introduce several fundamental concepts.

\begin{definition}\label{def5.1}
A shift operator $T_{\delta}$ is a linear operator acting on polynomials or
formal power series in the following manner
\begin{equation}\label{5.1}
T_{\delta}f(x) = f(x+\delta), \quad x \in \mathbb R, \quad \delta \in \mathbb R.
\end{equation}

For functions of several variables we introduce shift operators in the same
manner
\begin{eqnarray}
\lefteqn{T_{\delta_i}f(x_1, \dots x_{i-1}, x_i, x_{i+1}\dots x_n)}\nonumber\\
&&\hspace{1in} = f(x_1, \dots, x_{i-1}, x_i+\delta_i, x_{i+1}, \dots,
x_n).\label{5.2}
\end{eqnarray}
\end{definition}

In this section we restrict the exposition to the case of one real variable $x
\in \mathbb R$. The extension to $n$ variables and other fields is obvious.

\begin{definition}\label{def5.2}
An operator $U$ is called a ``delta operator'' if it satisfies the following
properties
\begin{enumerate}
\item[\textup{1)}] It is shift invariant;
\begin{equation}\label{5.3}
T_{\delta}U = UT_{\delta}, \quad \forall\delta \in \mathbb R
\end{equation}
\item[\textup{2)}] \begin{equation}\label{5.4}
Ux = c \neq 0, \quad c = \const
\end{equation}
\item[\textup{3)}] \begin{equation}\label{5.5}
Ua = 0, \quad \forall a
\end{equation}
\end{enumerate}
and the kernal of $U$ consists precisely of all constant.
\end{definition}

Important properties of delta operator are:
\begin{enumerate}
\item[1.] For every delta operator $U$ there exists a unique series of basic
polynomials $\{p_n(x)\}$ satisfying
\begin{equation}\label{5.6}
p_0(x) = 1 \quad p_n(0) = 0, \quad n \geq 1, \quad Up_n(x) = np_{n-1}(x).
\end{equation}

\item[2.] For every umbral operator $U$ there exists a conjugate operator
$\beta$, such that
\begin{equation}\label{5.7}
[U, x\beta] = 1.
\end{equation}
The operator $\beta$ satisfies
\begin{equation}\label{5.8}
\beta = (\stackrel{\prime}U)^{-1}, \quad \stackrel{\prime}U = [U, x].
\end{equation}
\end{enumerate}

The expression
\begin{equation}\label{5.9}
\stackrel{\prime}U \equiv U * x \equiv [U, x]
\end{equation}
is called the ``Pincherle derivative'' of $U$ \cite{ref52,ref53,ref54}.

For us the fundamental fact is that the pair of operators, $U$ and $x\beta$,
satisfies the Heisenberg relation (\ref{5.7}).

Before going further, let us give the two simplest possible examples.

\begin{example}\label{exam1}
The (continuous) derivative
\begin{equation}\label{5.10}
\begin{array}{c}
U = \partial_x, \quad \beta = 1\\
P_0 = 1, \quad P_1 = x, \quad \dots, P_n = x^n, \dots
\end{array}
\end{equation}
\end{example}

\begin{example}\label{exam2}
The right discrete derivative
\begin{equation}\label{5.11}
\begin{array}{c}
U = \Delta^+ = \frac{T-1}{\delta}, \quad \beta = T^{-1}\\
P_0 = 1, \quad P_1 = x, \quad P_2 = x(x-\delta)\\
P_n = x(x-\delta) \dots \big(x-(n-1)\delta\big).
\end{array}
\end{equation}
For any operator $U$ one can construct $\beta$ and the basic series will be
\begin{equation}\label{5.11b}
P_n = (x\beta)^n \cdot 1, \quad n \in \mathbb N.
\end{equation}
\end{example}

\subsection{Umbral Calculus and Linear Difference Equations}\label{subsec5.2}

First of all, let us consider a Lie algebra $L$ realized by vector fields
\begin{eqnarray}
&& X_a = f_a(x_1, \dots, x_n)\partial x_a\label{5.12}\\
&& [X_a, X_b] = C^c_{ab}X_c.\label{5.13} 
\end{eqnarray}
The Heisenberg relation (\ref{5.7}) allows us to realize the same abstract Lie
algebra by difference operators
\begin{equation}
X^D_a = f_a(x_1\beta_1, x_2\beta_2, \dots, x_n\beta_n)\Delta_{x_a}, \quad
[\Delta_{x_a}, x_a\beta_a] = 1, \quad a = 1, \dots n.
\end{equation}
As long as the functions $f_a$ are polynomials, or at least formal power series
in the variables $x_a$, the substitution
\begin{equation}\label{5.15}
x_a \to x_a\beta_a, \quad \partial_{x_a} \to \Delta_{x_a}
\end{equation}
preserves the commutation relations (\ref{5.13}).

We shall call the substitution (\ref{5.15}) and more generally any substitution
\begin{equation}\label{5.16}
\{U, \beta\} \leftrightarrow \{\tilde U, \tilde{\beta}\}
\end{equation}
an ``umbral correspondence''. This correspondence will also take the set of
basic polynomials related to $\{U, \beta\}$ into the set related to the pair
$\{\tilde u, \tilde{\beta}\}$.

We shall use two types of delta operators. The first is simply the derivative
$U = \partial_x$, for which we have $\beta = 1$. The second is a general
difference operator $U = \Delta$ that has $\partial_x$ is its continuous limit.
We put
\begin{equation}\label{5.17}
\Delta = \frac{1}{\delta}\sum^m_{k = l} a_k T^k_{\delta} \quad l, \quad
m \in \mathbb Z, \quad l < m
\end{equation}
where $a_k$ and $\delta$ are real constants and $T_{\delta}$ is a shift
operator as in eq.~(\ref{5.1}). Condition (\ref{5.3}) is satisfied. Condition
(\ref{5.5}) implies
\begin{equation}\label{5.18}
\sum^m_{k = l} a_k = 0.
\end{equation}
We also require that for $\delta \to 0$, we should have $\Delta \to
\partial_x$. This requires a further restriction on the coefficients $a_k$,
namely
\begin{equation}\label{5.19}
\sum^m_{k=l}  a_kk = 1.
\end{equation}
Then relation (\ref{5.4}) is also satisfied, with $c = 1$.

More generally, we have, for $\Delta$ as in (\ref{5.17})
\begin{eqnarray}
\Delta f(x) &=& \frac{1}{\delta} \sum^m_{k=l} a_kf(x+k\delta)\nonumber\\
&=& \frac{1}{\delta} \sum^{\infty}_{q=0} \frac{f^{(q)}(x)}{q!} \delta^q \sum^m
_{k=l} a_kk^q.\nonumber
\end{eqnarray}
We define
\begin{equation}\label{5.20}
\gamma_q = \sum^m_{k=l} a_kk^q \quad q \in \mathbb Z
\end{equation}
and thus
\begin{equation}
\Delta f(x) = \frac{df}{dx} + \sum^{\infty}_{q=2} \gamma_q \frac{f^{(q)}(x)}{q!}
\delta^{q-1}f.
\end{equation}

Thus $\Delta$ goes to the derivative at least to the order $\delta$. We can
also impose
\begin{equation}\label{5.21}
\gamma_q = 0, \quad q = 2, 3, \dots m - l.
\end{equation}
Then we have
\[
\Delta = \frac{d}{dx} + O(\delta^{m-1}).
\]

\begin{definition}\label{def5.3}
A difference operator of degree $m - l$ is a delta operator of the form
\begin{equation}\label{5.22}
U \equiv \Delta = \frac{1}{\delta} \sum^m_{k=l} a_k T^k_{\delta},
\end{equation}
where $a_k$ and $\delta$ are constants, $T_{\delta}$ is a shift operator and we
have
\begin{equation}\label{5.23}
\sum^m_{k=l} a_k = 0, \quad \sum^m_{k=l} a_kk = 1.
\end{equation}
\end{definition}

\noindent\emph{Comment}: $\tilde{\Delta} = T^j\Delta$ is a difference operator
of the same degree as $\Delta$.

\begin{theorem}\label{thm5.1}
The operator $\beta$ conjugate to $\Delta = (1/\delta) \sum^m_{k=l} a_k
T^k_{\delta}$ is
\begin{equation}\label{5.24}
\beta = \biggl(\sum^m_{k=l} a_kkT^k\biggr)^{-1}.
\end{equation}
\end{theorem}

\begin{proof}
$\beta = (\Delta')^{-1} = [\Delta, x]^{-1}$
\begin{eqnarray}
{}[\Delta, x] 
&=& \frac{1}{\delta}\biggl(\sum^m_{k=l}a_k(x+k\delta)T^k -
x\sum^m_{k=l}a_kT^k\biggr)\nonumber\\
&=& \sum a_kkT^k\nonumber.
\end{eqnarray}
\end{proof}

\noindent\emph{Examples}:
\begin{eqnarray}
\Delta^s &=& \frac{T-T^{-1}}{2\delta}, \quad \beta = \biggl(\frac{T+T^{-1}}
{2}\biggr)^{-1}\label{5.25}\\
\Delta^3 &=& - \frac{1}{6\delta}(T^2-6T+3+2T^{-1}), \quad \beta =
\biggl(-\frac{T^2-3T-T^{-1}}{3}\biggr)^{-1}\label{5.26}
\end{eqnarray}

\noindent\emph{Comment}:
\[
\Delta^s = \frac{\partial}{\partial x} + O(\delta^2) \quad \Delta^3 =
\frac{\partial}{\partial x} + O(\delta^3).
\]

Now let us apply the above considerations to the study of symmetries of linear
difference equations.

\begin{definition}\label{def5.4}
An umbral equation of order $n$ is an operator equation of the form
\begin{equation}\label{5.27}
\sum^n_{k=0} \hat a_k (x\beta) \Delta^k\hat f = \hat g 
\end{equation}
where $a_k(x\beta)$ and $\hat g(x\beta)$ are given formal power series in
$x\beta$ and $\hat f(x\beta)$ is the unknown operator function.
\end{definition}

For $\Delta = \partial_x$, $\beta = 1$ this is a differential equation. For
$\Delta$ as in (\ref{5.17}) eq.~(\ref{5.27}) is an operator equation. Applying
both sides of eq.~(\ref{5.27}) to 1 we get a difference equation. Its solution
is
\begin{equation}\label{5.28}
f(x) = \hat f(x\beta) \cdot 1.
\end{equation}
More generally, an umbral equation in $P$ variables is
\begin{eqnarray}
\sum^{n_1\dots n_p}_{k_1, \dots k_p} \hat a_{k_1\dots k_p}(x_1\beta_1, x_2
\beta_2, \dots, x_p\beta_p)\Delta^{k_1}_{\delta_1}, \dots
\Delta^{k_p}_{\delta_p}
\hat f(x_1\beta_1, \dots, x_p\beta_p)\nonumber\\
 = \hat g(x_1\beta_1\dots x_p\beta_p)\label{5.29},
\quad \sum^p_{i=1} n_i = n.
\end{eqnarray}

Example of an umbral equation
\begin{equation}\label{5.30}
\Delta\hat f = a\hat f, \quad a \neq 0.
\end{equation}
\begin{enumerate}
\item[(i)] Take $\Delta = \partial_x \Rightarrow f(x) = e^{ax}$

\item[(ii)] Take $\Delta = \Delta^+ = \frac{T-1}{\delta}, \quad \beta = T^{-1}$
\begin{equation}\label{5.31}
f(x+\delta) - f(x) = a\delta f(x).
\end{equation}
Take $f(x) = \lambda^x$:
\[
\lambda^{x+\delta} - \lambda^x  = a\delta\lambda^x \Rightarrow \lambda =
(1+a\delta)^{1/\delta}.
\]
We get a single ``umbral'' solution
\begin{equation}\label{5.32}
f_1(x) = (1+a\delta)^{x/\delta}.
\end{equation}
The umbral correspondence gives:
\begin{equation}\label{5.33}
f_2(x) = e^{axT^{-1}} \cdot 1.
\end{equation}

If we expand into power series, we obtain $f_1(x) = f_2(x)$, and of
course we have
\[
\lim_{\delta \to  0} f_{1, 2}(x) = e^{ax}.
\]

\item[(iii)] For comparison, take $\Delta = \Delta^s = (T-T^{-1})/2\delta$,
$\beta = [(T+T^{-1})/2]^{-1}$
\begin{equation}\label{5.34}
f(x+\delta) - f(x-\delta) = 2\delta af(x).
\end{equation}
Putting $f(x) = \lambda^x$ we get two values of $\lambda$ and
\begin{eqnarray}
f &=& A_1(a\delta + \sqrt{a^2\delta^2+1})^{x/\delta} + A_2(a\delta-\sqrt{a^2
\delta^2+1})^{x/\delta}\nonumber\\
&\equiv& A_1f_1 + A_2f_2\label{5.35}.
\end{eqnarray}
We have
\begin{equation}\label{5.36}
\lim_{\delta\to 0} f_1(x) = e^{ax},
\end{equation}
but the limit of $f_2(x)$ does not exist. The umbral correspondence yields
\[
f_u(x) = \exp\biggl[ax\biggl(\frac{T+T^{-1}}{2}\biggr)^{-1}\biggr] \cdot 1.
\]
Expanding into power series, we find $f_u = f_1$. The solution $f_2$ is a
nonumbral one.
\end{enumerate}

\begin{theorem}\label{thm5.2}
Let $\Delta$ be a difference operator of order $p$. Then the linear umbral
equation of order $n$ (\ref{5.27}) has $np$ linearly independent solutions, $n$
of them umbral ones.
\end{theorem}

There may be convergence problems for the formal series.

Consider the exponential
\begin{eqnarray}
\hat f(x) &=& e^{ax\beta}\nonumber\\
\beta &=& \biggl(\sum^m_{k=l} a_kkT^k\biggr)^{-1}\label{5.37}.
\end{eqnarray}
For $m - l \geq 3$, $\beta$ will involve infinitely many shifts i.e., each
term in the expansion (\ref{5.37}) could involve infinitely many shifts.
However
\begin{equation}\label{5.38}
P_n(x) = (x\beta)^n \cdot 1
\end{equation}
is a well defined polynomial. For a proof see \cite{ref33}.

Let us assume that we know the solution of an umbral equation for $\Delta =
\partial_x$ and it has the form
\begin{equation}\label{5.39}
f(x) = \sum^{\infty}_{n=0} \frac{f^{(n)}(0)}{n!} x^n.
\end{equation}
Then for any difference operator $\Delta$ there will exist a corresponding
umbral solution
\begin{equation}\label{5.40}
\hat f(x)1 = \sum^{\infty}_{n=0} \frac{f^{(n)}(0)}{n!} P_n(x),
\end{equation}
where $P_n(x) = (x\beta)^k \cdot 1$ are the basic polynomials corresponding to
$\Delta$.

\subsection{Symmetries of Linear Umbral Equations}\label{subsec5.3}

Let us consider a linear differential equation
\begin{equation}\label{5.41}
Lu = 0, \quad L = \sum_{k_1, \dots, k_p} a_{k_1, \dots, k_p}(x_1, \dots, x_p)
\frac{\partial^{k_1}}{\partial x^{k_1}_1} \dots \frac{\partial^{k_p}}{\partial
x^k_p}.
\end{equation}
The Lie point symmetries of eq.~(\ref{4.41}) can be realized by evolutionary
vector fields of the form
\begin{eqnarray}
\widehat X &=& Q(x_i, u, u_{, x_i})\partial_u,\nonumber\\
Q &=& \phi - \sum^p_{i=1} \xi_i u_{, x_i}\label{5.42}.
\end{eqnarray}
The following theorem holds for these symmetries.

\begin{theorem}\label{thm5.3}
All Lie point symmetries for an ODE of order $n \geq 3$, or a PDE of order $n
\geq 2$ are generated by evolutionary vector fields of the form
\textup{(\ref{5.42})} with the characteristic $Q$ satisfying
\begin{equation}\label{5.43}
Q = Xu + \chi(x_1, \dots, x_p),
\end{equation}
where $\chi$ is a solution of eq.~\textup(\ref{5.41}\textup) and $X$ is a linear
operator
\begin{equation}\label{5.44}
X = \sum^p_{i=1} \xi_i(x_1, \dots, x_p)\partial x_i
\end{equation}
satisfying
\begin{equation}\label{5.45}
{}[L, X] = \lambda(x_1, \dots, x_p)L,
\end{equation}
i.e.\ commuting with $L$ on the solution set of $L$. In
eq.~\textup(\ref{5.45}\textup) $\lambda$ is an arbitrary function.
\end{theorem}
For a proof we refer to the literature \cite{ref55}.

In other words, if the conditions of Theorem~\ref{thm5.3} apply, then all
symmetries of eq.~(\ref{4.41}) beyond those representing the linear
superposition principle, are generated by linear operators of the form
(\ref{5.44}), commuting with $L$ on the solution set of eq.~(\ref{5.41}).

Now let us turn to the umbral equation (\ref{5.29}) with $\hat g = 0$, i.e.
\begin{equation}\label{5.46}
\sum_{k_1\dots k_p} \hat a_{k_1\dots k_p}(x_1\beta_1, \dots x_p\beta_p)
\Delta^{k_1}_{\delta_1} \dots \Delta^{k_p}_{\delta_p} \hat u(x_1\beta_1, \dots,
x_p\beta_p) = 0.
\end{equation}

We shall realize the symmetries of eq.~(\ref{5.46}) by evolutionary vector
fields of the form
\begin{equation}\label{5.47}
v_E = Q_D\partial_u, \quad Q_D = \phi_D - \sum^p_{i=1}\xi_{D, i} \Delta_iu 
\end{equation}
where $\phi_D$ and $\xi_{D, i}$ are functions of $x_i\beta_i$ and $u$. The
prolongation of $v_E$ will also act on the discrete derivatives
$\Delta^{k_i}_{\delta_i}u$. We are now considering transformations on a fixed
(nontransforming) lattice. In the evolutionary formalism the transformed
variables satisfy
\begin{eqnarray}
\tilde x_k\tilde{\beta}_k &=& x_k\beta_k, \quad \tilde{\beta}_k =
\beta_k\nonumber\\
\tilde u(\tilde x_k\tilde{\beta}_k) &=& u(x_k\beta_k) + \lambda
Q_D, \quad |\lambda| \ll 1\label{5.48}
\end{eqnarray}
and we request that $\tilde{\psi}$ be a solution whenever $\psi$ is one.

The transformation of the discrete derivatives is given by
\begin{eqnarray}
\Delta_{\tilde x_k}\tilde u &=& \Delta_{x_k}u + \lambda\Delta_{x_k}Q
\nonumber\\
\Delta_{\tilde x_k\tilde x_k}\tilde u &=& \Delta_{x_kx_k}u +
\lambda\Delta_{x_kx_k} Q\label{5.49}
\end{eqnarray}
etc., where $\Delta_{x_k}$ are discrete total derivatives.

In terms of the vector field (\ref{5.57}) we have
\begin{eqnarray}
\pr v_E & = & Q_D\partial_u + Q^{x_i}_D\partial_{\Delta_iu} + Q^{x_ix_k}_D
\partial_{\Delta_i\Delta_k u} + \dots\nonumber\\
Q^{x_i}_D & = & \Delta_i Q_D, \quad Q^{x_ix_k}_D =
\Delta_i\Delta_kQ_D\label{5.50}
\end{eqnarray}
(we have put $\Delta_{\delta_i} \equiv \Delta_{x_i} \equiv \Delta_i$).

As in the continuous continuous case, we obtain determining equations by
requiring
\begin{equation}\label{5.51}
\pr v_E(L_D\hat u)\mid_{L_D\hat u} = 0
\end{equation}
where $L_D \hat u$ is the left hand side of eq.~(\ref{5.46}).

The determining equations will be an umbral version of the determining equations
in the continuous case, i.e.\ are obtained by the umbral correspondence
$\partial_{x_i} \to \Delta_i$, $x_i \to x_i\beta_i$.

The symmetries of the umbral equation (\ref{5.46}) will hence have the form
(\ref{5.47}) with
\begin{equation}\label{5.52}
Q_D = X_Du + \chi(x_1\beta_1, \dots x_p\beta_p)
\end{equation}
where $X_D$ is a difference operator commuting with $L_D$ on the solutions of
eq.~(\ref{5.46}). Moreover, $X_D$ is obtained from $X$ by the umbral
correspondence.

We shall call such symmetries ``generalized point symmetries''. Because of the
presence of the operators $\beta_i$ they are not really point symmetries. In
the continuous limit they become point symmetries.

Let us now consider some examples.

\subsection{Example of the Discrete Heat Equation}\label{subsec5.4}

The (continuous) linear heat equation in $(1+1)$ dimensions is
\begin{equation}\label{5.53}
u_t - u_{xx} = 0.
\end{equation}
Its symmetry group is of course well-known. Factoring out the infinite
dimensional pseudo-group corresponding to the linear superposition principle we
have a 6 dimensional symmetry group. We write its Lie algebra in evolutionary
form as
\begin{eqnarray}
P_o &=& u_t\partial_u, \quad P_1 = u_x\partial_u, \quad W = u\partial_u
\nonumber\\
B &=& (2tu_x+xu)\partial_u, \quad D = \biggl(2tu_t+xu_x+\frac 12 u\biggr)
\partial_u\label{5.54}\\
K &=& \biggl[t^2u_t+txu_x+\frac 14(x^2+2t)u\biggr]\partial_u\nonumber
\end{eqnarray}
where $P_o$, $P_1$, $B$, $D$, $K$ and $W$ generate time and space translations,
Galilei boosts, dilations ``expansions'' and the multiplication of $u$ by a
constant, respectively.

A very natural discretization of eq.~(\ref{5.53}) is the discrete heat equation
\begin{equation}\label{5.55}
\Delta_tu - (\Delta_x)^2u = 0,
\end{equation}
where $\Delta_t$ and $\Delta_x$ are each one of the difference operators
considered in Section~\ref{subsec5.2}. We use the corresponding conjugate
operators $\beta_t$ and $\beta_x$, respectively. The umbral correspondence
gives us the symmetry algebra of eq.~(\ref{5.55}), starting from the algebra
(\ref{5.54}). Namely, we have
\begin{eqnarray}
P^D_0 &=& (\Delta_tu)\partial_u, \quad P^D_1 = (\Delta_xu)\partial_u, \quad W^D
= u\partial_u\nonumber\\
B^D &=& [2(t\beta_t)\Delta_xu+(x\beta_x)u]\partial_u\nonumber\\
D^D &=& \biggl[2t\beta_t\Delta_tu+x\beta_x\Delta_xu+\frac 12u\biggr]\partial_u
\label{5.56}\\
K^D &=& \biggl[(t\beta_t)^2\Delta_tu+(t\beta_t)(x\beta_x)\Delta_xu+\frac
14\big((x\beta_x)^2+2t\beta_t\big)u\biggr]\partial u\nonumber.
\end{eqnarray}

In particular we can choose both $\Delta_t$ and $\Delta_x$ to be right
derivatives
\begin{equation}\label{5.57}
\Delta_t =\frac{T_t-1}{\delta_t}, \quad \beta_t = T^{-1}_t, \quad \Delta_x =
\frac{T_x-1}{\delta_x}, \quad \beta_x = T^{-1}_x,
\end{equation}

The characteristic $Q_k$ of the element $K^D$ then is
\begin{eqnarray}
\lefteqn{Q_K = X_Ku, \quad X_K = (t^2-\delta_tt)T^{-2}_t \Delta_t +
txT^{-1}_xT^{-1}_t\Delta_x}\nonumber\\
&\hspace{2in}& + \frac 14[(x^2-\delta x)T^{-2}_x + 2tT^{-1}_t],\label{5.58}
\end{eqnarray}
so it is not a point transformation: it involves $u$ evaluated at several
points. Each of the basis elements (\ref{5.56}) (or any linear combination of
them) provides a flow commuting with eq.~(\ref{5.55}):
\begin{equation}\label{5.59}
u_{\lambda} = Xu.
\end{equation}
Equations (\ref{5.55}) and (\ref{5.59}) can be solved simultaneously and this
will provide a difference analog of the separation of variables in PDEs and a
tool for studying new types of special functions.

\subsection{The Discrete Burgers Equation and its Symmetries}\label{subsec5.5}

\subsubsection{The Continuous Case}\label{subsubsec4.5.1}

The Burgers equation
\begin{equation}\label{5.60}
u_t = u_{xx} + 2uu_x
\end{equation}
is the simplest equation that combines nonlinearity and dissipative effects. It
is also the prototype of an equation linearizable by a coordinate
transformation $C$-linearizable in Calogero's terminology \cite{ref59}.

We put $u = v_x$ and obtain the potential Burgers equation for $v$:
\begin{equation}\label{5.61}
v_t = v_{xx} + v^2_x.
\end{equation}
Putting $w = e^v$ we find
\begin{equation}\label{5.62}
w_t = w_{xx}.
\end{equation}
In other words, the usual Burgers equation~(\ref{5.60}) is linearized (into the
heat equation (\ref{5.62}) by the Cole-Hopf transformation
\begin{equation}\label{5.63}
u = \frac{w_x}{w}
\end{equation}
(which is not a point transformation).

One possible way of viewing the Cole-Hopf transformation is that it provides a
Lax pair for the Burgers equation:
\begin{equation}\label{5.64}
w_t = w_{xx}, \quad w_x = uw.
\end{equation}
Putting
\[
w_t = Aw, \quad w_x = Bw, \quad A = u_x + u^2, \quad B = u
\]
we obtain the Burgers equation as a compatibility condition
\begin{equation}\label{5.65}
A_x - B_t + [A, B] = 0.
\end{equation}

Our aim is to discretize the Burgers equation in such a way as to preserve its
linearizability and also its five-dimensional Lie point symmetry algebra. We
already know the symmetries of the discrete heat equation and we will use them
to obtain the symmetry algebra of the discrete Burgers equation. This will be
an indirect application of umbral calculus to a nonlinear equation.

\subsubsection{The Discrete Burgers Equation as a Compatibility Condition}
\label{subsubsec5.5.2}

Let us write a discrete version of the pair (\ref{5.64}) in the form:
\begin{equation}\label{5.66}
\Delta_t\phi = \Delta_{xx}\phi, \quad \Delta_x\phi = u\phi
\end{equation}
where we take
\begin{equation}\label{5.67}
\Delta_t = \frac{T_t-1}{\delta_t}, \quad \Delta_x = \frac{T_x-1}{\delta_x}.
\end{equation}
The pair (\ref{5.66}) can be rewritten as
\begin{equation}\label{5.68}
\Delta_t\phi = (\Delta_xu+uT_xu)\phi, \quad \Delta_x\phi = u\phi.
\end{equation}

We have used the Leibnitz rule appropriate for the discrete derivative
$\Delta_x$ of (\ref{5.67}), namely
\begin{equation}\label{5.69}
\Delta_xfg = f(x)\Delta_xg + (T_xg)\Delta_xf.
\end{equation}
Compatibility of eq.~(\ref{5.68}), i.e. $\Delta_x\Delta_t\phi =
\Delta_t\Delta_x\phi$ yields the discrete Burgers equation
\begin{equation}\label{5.70}
\Delta_tu = \frac{1+\delta_xu}{1+\delta_t[\Delta_x\Delta_xu+uT_xu]}
\Delta_x(\Delta_xu+uT_xu).
\end{equation}
In the continuous limit $\Delta_t \to \partial/\partial t$, $\Delta_x  \to
\partial/\partial x$, $T_x \to 1, \delta_x = 0$, $\delta_t = 0$ we
reobtain the Burgers equation (\ref{5.60}) \cite{ref56,ref32}. This is not a 
``naive'' discretization like
\begin{equation}\label{5.71}
\Delta_tu  = (\Delta_x)^2 u + 2u \Delta_xu
\end{equation}
which would loose all integrability properties.

\subsubsection{Symmetries of the discrete Burgers
Equation}\label{subsubsec5.5.3}

We are looking for ``generalized point symmetries'' on a fixed lattice. We
write them in evolutionary form
\begin{equation}\label{5.72}
X_e = Q(x, t, T^a_xT^b_tu, T^c_xT^d_t\Delta_xu, T^e_xT^f_t\Delta_tu, \dots)
\partial_u
\end{equation}
and each symmetry will provide a commuting flow
\[
u_{\lambda} = Q.
\]
We shall use the Cole-Hopf transformation to transform the symmetry algebra of
the discrete heat equation into that of the discrete Burgers equation.

All the symmetries of the discrete heat equation given in eq.~(\ref{5.56}) can
be written as
\begin{equation}\label{5.73}
\phi_{\lambda} = S\phi, \quad S = S(x, t, \phi, T_x, T_x \dots)
\end{equation}
where $S$ is a linear operator (the same is true for any linear difference
equation).

For the discrete heat equation
\begin{equation}\label{5.74}
\Delta_t\phi - (\Delta_x)^2\phi = 0
\end{equation}
with $\Delta_t$ and $\Delta_x$ as in eq.~(\ref{5.67}) we rewrite the flows
corresponding to eq.~(\ref{5.56}) as
\begin{eqnarray}
\phi_{\lambda_1} &=& \Delta_t\phi, \quad \phi_{\lambda_2} = \Delta_x\phi, \quad
\phi_{\lambda_3} = \biggl[2tT^{-1}_t\Delta_x+xT^{-1}_x+\frac 12\delta_xT^{-1}_x
\biggr]\phi\nonumber\\
\phi_{\lambda_4} &=& \biggl[2tT^{-1}_t\Delta_t+xT^{-1}_x\Delta_x+\frac
12\biggr]\phi\nonumber\\
\phi_{\lambda_5} &=& \biggl[t^2T^{-2}_t\Delta_t+txT^{-1}_tT^{-1}_x\Delta_x+\frac
14x^2T^{-2}_x\nonumber\\
&& \hspace{0.75in} + t\biggl(T^{-2}_t-\frac 12T^{-1}_tT^{-1}_x\biggr)-\frac{1}
{16}\delta^2_xT^{-2}_x\biggr]\phi\label{5.75}\\
\phi_{\lambda_6} &=& \phi.\nonumber
\end{eqnarray}

Let us first prove a general result.

\begin{theorem}\label{thm5.4}
Let eq.~\textup{(\ref{5.73})} represent a symmetry of the discrete heat
equation \textup{(\ref{5.74})}. Then the same linear operator $S$ provides
a symmetry of the discrete Burgers equation \textup{(\ref{5.70})}, the
flow of which is given by
\begin{equation}\label{5.76}
u_{\lambda} = (1+\delta_xu)\Delta_x\biggl(\frac{S\phi}{\phi}\biggr),
\end{equation}
where $(S\phi)/\phi$ can be expressed in terms of $u(x, t)$.
\end{theorem}

\begin{proof}
We request that eq.~(\ref{5.73}) and the Cole-Hopf transformation in
eq.~(\ref{5.66}) be compatible
\begin{equation}\label{5.77}
\frac{\partial}{\partial\lambda}(\Delta_x\phi) = \Delta_x\phi_{\lambda}.
\end{equation}
>From here we obtain
\begin{equation}\label{5.78}
u_{\lambda} = \frac{\Delta_x(S\phi)-uS\phi}{\phi}.
\end{equation}
A direct calculation yields
\begin{equation}\label{5.79}
\Delta_x\biggl(\frac{S\phi}{\phi}\biggr) = \frac{1}{T_x\phi}[\Delta_x(S\phi)-u
(S\phi)]
\end{equation}
and (\ref{5.76}) follows. It is still necessary to show that $S\phi/\phi$
depends only on $u(x, t)$. The expressions for $S\phi$ can be read off from
eq.~(\ref{5.75}). From there we see that all expressions involved can be
expressed in terms of $u(x, t)$ and its shifted values (using the Cole-Hopf
transformation again). Indeed, we have
\begin{eqnarray}
\Delta_x\phi &=& u\phi, \quad \Delta_t\phi = v\phi\nonumber\\
T_x\phi &=& (1+\delta_xu)\phi, \quad T_t\phi = (1+\delta_tv)\phi\label{5.80}\\
T^{-1}_x\phi &=& \biggl(T^{-1}_x\frac{1}{1+\delta_xu}\biggr)\phi, \quad T^{-1}
_t\phi  = \biggl(T^{-1}_t\frac{1}{1+\delta_tv}\biggr)\phi,\nonumber
\end{eqnarray}
where we define
\begin{equation}\label{5.81}
v = \Delta_xu + uT_xu.
\end{equation}
Explicitly, eq.~(\ref{5.76}) maps the 6 dimensional symmetry algebra of the
discrete heat equation into the 5 dimensional Lie algebra of the discrete
Burgers equation. The corresponding flows are
\begin{eqnarray}
u_{\lambda_1}  &=& (1+\delta_tv)\Delta_tu\nonumber\\
u_{\lambda_2} &=& (1+\delta_xu)\Delta_xu\nonumber\\
u_{\lambda_3} &=& (1+\delta_xu)\Delta_x\biggl[2tT^{-1}_t\frac{u}{1+\delta_tv}
 + \biggl(x+\frac 12 - \delta_x\biggr)T^{-1}_x\frac{1}{1+\delta_xu}\biggr]
\nonumber\\
u_{\lambda_4} &=& (1+\delta_xu)\Delta_x\biggl[2tT^{-1}_t\frac{v}{1+\delta_tv}
 + xT^{-1}_x \frac{u}{1+\delta_xu}
 - \frac 12 T^{-1}_x\frac{1}{1+\delta_xu}\biggr]\nonumber\\
u_{\lambda_5} &=& (1+\delta_xu)\Delta_x\biggl[t^2T^{-1}_t\biggl(
\frac{1}{1+\delta_tv} T^{-1}_t\frac{v}{1+\delta_tv}\biggr)\label{5.82}\\
&&\hspace{1.5in} + txT^{-1}\biggl(\frac{1}{1+\delta_xu}T^{-1}_t
\frac{u}{1+\delta_tv}\biggr)\nonumber\\
&&\hspace{1.5in} + \frac 14\biggl(x^2-\frac{\delta^2_x}{4}\biggr)T^{-1}_x
\biggl(\frac{1}{1+\delta_xu}T^{-1}_x\frac{1}{1+\delta_xu}\biggr)\nonumber\\
&&\hspace{1.5in} + tT^{-1}_t
\biggl(\frac{1}{1+\delta_tv}T^{-1}_t\frac{1}{1+\delta_tv}\biggr)\nonumber\\
&&\hspace{1.5in} -\frac 12
tT^{-1}_x\biggl(\frac{1}{1+\delta_xu}T^{-1}_t\frac{1}{1+\delta_tv}\biggr)
\biggr]\nonumber\\
u_{\lambda_6} &=& 0.\nonumber
\end{eqnarray}
\end{proof}

\subsubsection{Symmetry Reduction for the Discrete Burgers Equation}
\label{subsubsec5.5.4}

Symmetry reduction for continuous Burger equation: we add a compatible equation
to the Burgers equation
\begin{eqnarray}
u_t &=& u_{xx} + 2uu_x\nonumber\\
u_{\lambda} &=& Q(x, t, u, u_{x, t}) = 0\label{5.83}
\end{eqnarray}
and solve the two equations simultaneously. Example: time translations.
\begin{equation}\label{5.84}
u_{\lambda} = u_t = 0.
\end{equation}
Then $u = u(x)$ and
\begin{equation}\label{5.85}
u_{xx} + 2uu_x = 0 \Rightarrow u_x + u^2 = K.
\end{equation}
>From here, we obtain three types of solutions
\begin{equation}\label{5.86}
u = \frac 1x, \quad u = k \arctanh kx, \quad u = k \arctan kx.
\end{equation}
Symmetry reduction in the discrete case. All flows have the form (\ref{5.76}).
The condition $u_{\lambda} = 0$ hence implies
\begin{equation}\label{5.87}
S\phi = K(t)\phi,
\end{equation}
where $K(t)$ is an arbitrary function. This equation must be solved together
with the discrete Burgers equation in order to obtain group invariant solutions.

Let us here consider just one example, namely that of time translations, the
first equation in (\ref{5.82}). Eq.~(\ref{5.87}) reduces to
\begin{equation}\label{5.88}
\Delta_t\phi = K(t)\phi,
\end{equation}
i.e.
\begin{equation}\label{5.89}
v = \Delta_xu + uT_xu = K(t).
\end{equation}
We rewrite the Burgers equation as
\begin{equation}\label{5.90}
\Delta_tu = \frac{1+\delta_xu}{1+\delta_tv} \Delta_xv \quad v \equiv \Delta_xu
+ uT_xu.
\end{equation}
However, from (\ref{5.89}) we have $v = K(t)$ and hence $\Delta_tu = 0$, $K(t)
= K_o = \const$. Since $\phi$ satisfies the heat equation, we can rewrite
(\ref{5.88}) as
\begin{equation}\label{5.91}
\Delta_{xx}\phi = K_o\phi.
\end{equation}
The general solution of (\ref{5.91}) is obtained for $K_o \neq 0$ by putting
$\phi = a^x$ and solving (\ref{5.91}) for $a$. We find
\begin{equation}\label{5.92}
\phi = c_1(1+\sqrt{K_0}\delta_x)^{x/\delta_x} +
c_2(1-\sqrt{K_0}\delta_x)^{x/\delta_x}.
\end{equation}
For $K_o = 0$ we have
\begin{equation}\label{5.93}
\phi = c_1 + c_2x.
\end{equation}
Solutions of the discrete Burgers equation are obtained via the Cole-Hopf
transformation
\begin{equation}\label{5.94}
u = \frac{\Delta_x\phi}{\phi}.
\end{equation}
The  same procedure can be followed for all other symmetries. We obtain linear
second order difference equations for $\phi$ involving one variable only.
However, the equations have variable coefficients and are hard to solve. They
can be reexpressed as equations for $u(x, t)$, again involving only one
independent variable. Thus, a reduction takes place, but it is not easy to
solve the reduced equations explicitly.

For instance, Galilei invariant solutions of the discrete Burgers equation must
satisfy the ordinary difference equation
\begin{equation}\label{5.95}
\begin{array}{c}
2tT_xu + x - K(t) + 2t\delta_xuT_xu + \delta_t\biggl(\frac 72T_xu + \frac 72
\delta_xuT_xu\\
+ xuT_xu + x\Delta_xu - \frac 32u\biggr) + \frac 32\delta_x - K(t)[\delta_xu +
\delta_t(T_x\Delta_xu\\
+ uT^2_xu -uT_xu) + T_xuT^2_xu + \delta_xuT_xuT^2_xu] = 0
\end{array}
\end{equation}
where $t$ figures as a parameter.

\section*{Acknowledgements}

I thank the organizers of the CIMPA School for giving me the opportunity to present these lectures. Special thanks go to the Tamizhmani family for making my visit to Pondicherry both pleasant and memorable.

The research reported upon in these lectures was partly supported by research grants from NSERC of Canada, FQRNT du Qu\'ebec and the NATO collaborative grant n.~PST.CLG.978431.


\printindex
\end{document}